\documentclass{vldb}

\usepackage{scrextend, MnSymbol}
\usepackage[hyphens]{url}

\usepackage[hidelinks]{hyperref}

\usepackage{breakurl}

\newif\iftechreport
\techreporttrue

\usepackage{times, graphicx, xcolor, multicol, minipage, subfig, tikz, amsthm, float}
\usepackage{balance}  
\newcommand{\subparagraph}{}

\usepackage{xspace, titlesec}

\newcommand{\figfactor}{1.0}

\usepackage{algorithm}
\usepackage[noend]{algpseudocode}
\newcommand{\commentt}[1]{{\small\texttt{#1}}}

\theoremstyle{definition}

\newtheorem{definition}{Definition}

\makeatletter
\def\th@plain{%
  \thm@notefont{}
  \itshape 
}
\def\th@definition{%
  \thm@notefont{}
  \normalfont 
}
\makeatother

\usetikzlibrary{positioning, snakes, shapes}

\tikzset{>=latex}

\titleformat*{\section}{\fontsize{12pt}{12}\selectfont\bfseries}
\titleformat*{\subsection}{\fontsize{10pt}{11}\selectfont\bfseries}
\titleformat*{\subsubsection}{\fontsize{10pt}{11}\itshape}

\titlespacing{\section}{1ex}{1ex}{0.25ex}

\titlespacing{\subsection}{1ex}{0.5ex}{0.25ex}
\titlespacing{\subsubsection}{1ex}{0.25ex}{0.25ex}

\let\footnotesize\small

\newcommand*{\DashedArrow}[1][]{\mathbin{\tikz [baseline=-.25ex,-latex, dashed,#1] \draw [#1] (0pt,0.5ex) -- (2em,0.5ex);}}%

\newenvironment{myenumerate}
{

   \vspace{-.5em}
   \newcounter{qcounter}
    \begin{list}{\arabic{qcounter}.~}{\usecounter{qcounter}\leftmargin=1em}
        \setlength{\topsep}{0em}
        \setlength{\parskip}{0pt}
        \setlength{\partopsep}{0pt}
        \setlength{\parsep}{0pt}         
        \setlength{\itemsep}{.25em} 
        \setlength{\itemindent}{0em}
}
{
    \end{list} 
    \vspace{-.5em}
}

\newenvironment{introenumerate}
{

   \vspace{-.5em}
   \newcounter{qdcounter}
    \begin{list}{\arabic{qdcounter}.~}{\usecounter{qdcounter}\leftmargin=1em}
        \setlength{\topsep}{0em}
        \setlength{\parskip}{0pt}
        \setlength{\partopsep}{0pt}
        \setlength{\parsep}{0pt}         
        \setlength{\itemsep}{.25em} 
        \setlength{\itemindent}{0em}
}
{
    \end{list} 
    \vspace{-.5em}
}

\begin{document}

\setlength{\abovedisplayskip}{4pt}
\setlength{\abovedisplayshortskip}{4pt}
\setlength{\belowdisplayskip}{4pt}
\setlength{\belowdisplayshortskip}{4pt}

\pagestyle{empty}


\title{Highly Available Transactions: Virtues and Limitations\iftechreport\\(Extended Version)\fi}
{\author{Peter Bailis, Aaron Davidson, Alan Fekete{\fontsize{12}{14}$^\dagger$}, Ali Ghodsi, Joseph M. Hellerstein, Ion Stoica \\[1mm]{\affaddr{UC Berkeley and {\fontsize{12}{14}$^\dagger$}University of Sydney}}}}
\maketitle

\begin{abstract}
\noindent To minimize network latency and remain online during server
failures and network partitions, many modern distributed data storage
systems eschew transactional functionality, which provides strong
semantic guarantees for groups of multiple operations over multiple
data items. In this work, we consider the problem of providing Highly
Available Transactions (HATs): transactional guarantees that do not
suffer unavailability during system partitions or incur high network
latency.  We introduce a taxonomy of highly available systems and
analyze existing ACID isolation and distributed data consistency
guarantees to identify which can and cannot be achieved in HAT
systems. This unifies the literature on weak transactional isolation,
replica consistency, and highly available systems. We analytically and
experimentally quantify the availability and performance benefits of
HATs---often two to three orders of magnitude over wide-area
networks---and discuss their necessary semantic compromises.
\end{abstract}

\section{Introduction}

The last decade has seen a shift in the design of popular large-scale
database systems, from the use of transactional
RDBMSs~\cite{bernstein-book, gray-isolation, gray-virtues} to the
widespread adoption of loosely consistent distributed key-value
stores~\cite{bigtable, pnuts, dynamo}. Core to this shift was the 2000
introduction of Brewer's CAP Theorem, which stated that a highly
available system cannot provide ``strong'' consistency guarantees in
the presence of network partitions~\cite{brewer-slides}. As formally
proven~\cite{gilbert-cap}, the CAP Theorem pertains to a data
consistency model called linearizability, or the ability to read the
most recent write to a data item that is replicated across
servers~\cite{herlihy-art}. However, despite its narrow scope, the CAP
Theorem is often misconstrued as a broad result regarding the ability
to provide ACID database properties with high
availability~\cite{hat-hotos,brewer-slides, foundation-article}; this
misunderstanding has led to substantial confusion regarding replica
consistency, transactional isolation, and high availability. The
recent resurgence of transactional systems suggests that programmers
value transactional semantics, but most existing transactional data
stores do not provide availability in the presence of
partitions~\cite{orleans,foundation-article, hstore,spanner,eiger,
  walter,calvin}.

Indeed, serializable transactions---the gold standard of traditional
ACID databases---are not achievable with high availability in the
presence of network partitions~\cite{davidson-survey}. However, database
systems have a long tradition of providing weaker isolation and
consistency guarantees~\cite{adya, ansicritique, gray-virtues,
  gray-isolation, kemme-thesis}. Today's ACID and NewSQL databases
often employ weak isolation models due to concurrency and performance
benefits; weak isolation is overwhelmingly the default setting in
these stores and is often the only option offered
(Section~\ref{sec:modernacid}). While weak isolation levels do not
provide serializability for general-purpose transactions, they are
apparently strong enough to deliver acceptable behavior to many
application programmers and are substantially stronger than the
semantics provided by current highly available systems. This raises a
natural question: which semantics can be provided with high
availability?

To date, the relationship between ACID semantics and high
availability has not been well explored. We have a strong
understanding of weak isolation in the single-server context from
which it originated~\cite{adya, ansicritique, gray-isolation} and many
papers offer techniques for providing distributed
serializability~\cite{bernstein-book, spanner, daudjee-session,
  hstore, calvin} or snapshot
isolation~\cite{kemme-thesis,walter}. Additionally, the distributed computing and parallel
hardware literature contains many consistency models for single
operations on replicated objects~\cite{pnuts, herlihy-art, eiger, cac,
  sessionguarantees}. However, the literature lends few clues for
providing semantic guarantees for multiple operations operating on
multiple data items in a highly available distributed environment.

Our main contributions in this paper are as follows. We relate the
many previously proposed database isolation and data consistency
models to the goal of high availability, which guarantees a response
from each non-failing server in the presence of arbitrary network
partitions between them.  We classify which among the wide array of
models are achievable with high availability, denoting them as {\em
  Highly Available Transactions} (HATs). In doing so, we demonstrate
that although many implementations of HAT semantics are not highly
available, this is an artifact of the implementations rather than an
inherent property of the semantics. Our investigation shows that,
besides serializability, Snapshot Isolation and Repeatable Read
isolation are not HAT-compliant, while most other isolation levels are
achievable with high availability. We also demonstrate that many weak
replica consistency models from distributed systems are both
HAT-compliant and simultaneously achievable with several ACID
properties.

Our investigation is based on both impossibility results and several
constructive, proof-of-concept algorithms. For example, Snapshot
Isolation and Repeatable Read isolation are not HAT-compliant because
they require detecting conflicts between concurrent updates (as needed
for preventing Lost Updates or Write Skew phenomena), which we show is
unavailable. However, Read Committed isolation, transactional
atomicity (Section~\ref{sec:ta}), and many other consistency models
from database and distributed systems are achievable via algorithms
that rely on multi-versioning and limited client-side caching. For
several guarantees, such as causal consistency with phantom prevention
and ANSI Repeatable Read, we consider a modified form of high
availability in which clients ``stick to'' (i.e., have affinity with)
at least one server---a property which is often implicit in the
distributed systems literature~\cite{herlihy-art, eiger, cac} but
which requires explicit consideration in a client-server replicated
database context. This sticky availability is widely
employed~\cite{eiger, vogels-defs} but is a less restrictive model
(and therefore more easily achievable) than traditional high
availability.

At a high level, the virtues of HATs are guaranteed responses from any
replica, low latency, and a range of semantic guarantees including
several whose usefulness is widely accepted such as Read
Committed. However, highly available systems are fundamentally unable
to prevent concurrent updates to shared data items and cannot provide
recency guarantees for reads. To understand when these virtues and
limitations are relevant in practice, we survey both practitioner
accounts and academic literature, perform experimental analysis on
modern cloud infrastructure, and analyze representative applications
for their semantic requirements. Our experiences with a HAT prototype
running across multiple geo-replicated datacenters indicate that HATs
offer a one to three order of magnitude latency decrease compared to
traditional distributed serializability protocols, and they can
provide acceptable semantics for a wide range of programs, especially
those with monotonic logic and commutative updates~\cite{calm,
  crdt}. HAT systems can also enforce arbitrary foreign key
constraints for multi-item updates and, in some cases, provide limited
uniqueness guarantees. However, HATs can fall short for applications
with concurrency-sensitive operations, requiring unavailable,
synchronous coordination.

Finally, we recognize that the large variety of ACID isolation levels
and distributed consistency models (and therefore those in our
taxonomy) can be confusing; the subtle distinctions between models may
appear to be of academic concern. Accordingly, we offer the following
pragmatic takeaways:
\begin{introenumerate}
\item The default (and sometimes strongest) configurations of most
  widely deployed database systems expose a range of anomalies that
  can compromise application-level consistency.

\item Many of these ``weak isolation'' models are achievable without
  sacrificing high availability if implemented correctly. However,
  none of the achievable models prevents concurrent modifications.

\item In addition to providing a guaranteed response and horizontal
  scale-out, these highly available HAT models allow one to three
  order of magnitude lower latencies on current infrastructure.

\item For correct behavior, applications may require a combination of
  HAT and (ideally sparing use of) non-HAT isolation levels; future
  database designers should plan accordingly.
\end{introenumerate}




\section{Why High Availability?}
\label{sec:motivation}

Why does high availability matter? Peter Deutsch starts his classic
list of ``Fallacies of Distributed Computing'' with two concerns
fundamental to distributed database systems: ``\textit{1.)}  The
network is reliable. \textit{2.)} Latency is
zero''~\cite{fallacies-deutsch}. In a distributed setting, network
failures may prevent database servers from communicating, and, in the
absence of failures, communication is slowed by factors like physical
distance, network congestion, and routing. As we will see
(Section~\ref{sec:availability}), highly available system designs
mitigate the effects of network \textit{partitions} and
\textit{latency}. In this section, we draw on a range of evidence that
indicates that partitions occur with frequency in real-world
deployments and latencies between datacenters are substantial, often
on the order of several hundreds of milliseconds.

\subsection{Network Partitions at Scale}

According to James Hamilton, Vice President and Distinguished Engineer
on the Amazon Web Services team, ``network partitions should be rare
but net gear continues to cause more issues than it
should''~\cite{hamilton-partitions}. Anecdotal evidence confirms
Hamilton's assertion. In April 2011, a network misconfiguration led to
a twelve-hour series of outages across the Amazon EC2 and RDS
services~\cite{amazon-netpartition}. Subsequent misconfigurations and
partial failures such as another EC2 outage in October 2012 have led
to full site disruptions for popular web services like Reddit,
Foursquare, and Heroku~\cite{ec2-downsites}. At global scale, hardware
failures---like the 2011 outages in Internet backbones in North
America and Europe due a router bug~\cite{juniper-partition}---and
misconfigurations like the BGP faults in 2008~\cite{pakistan-youtube}
and 2010~\cite{research-experiment-partition} can cause widespread
partitioning behavior.

Many of our discussions with practitioners---especially those
operating on public cloud infrastructure---as well as reports from
large-scale operators like Google~\cite{dean-keynote} confirm that
partition management is an important consideration for service
operators today. System designs that do not account for partition
behavior may prove difficult to operate at scale: for example, less
than one year after its announcement, Yahoo!'s PNUTS developers
explicitly added support for weaker, highly available operation. The
engineers explained that ``strict adherence [to strong consistency]
leads to difficult situations under network partitioning or server
failures...in many circumstances, applications need a relaxed
approach''~\cite{pnuts-update}.

Several recent studies rigorously quantify partition behavior. A 2011
study of several Microsoft datacenters observed over 13,300 network
failures with end-user impact, with an estimated median 59,000 packets
lost per failure. The study found a mean of 40.8 network link failures
per day (95th percentile: 136), with a median time to repair of around
five minutes (and up to one week). Perhaps surprisingly, provisioning
redundant networks only reduces impact of failures by up to 40\%,
meaning network providers cannot easily curtail partition
behavior~\cite{sigcomm-dc}. A 2010 study of over 200 wide-area routers
found an average of 16.2--302.0 failures per link per year with an
average annual downtime of 24--497 minutes per link per year (95th
percentile at least 34 hours)~\cite{sigcomm-wan}. In HP's managed
enterprise networks, WAN, LAN, and connectivity problems account for
28.1\% of all customer support tickets while 39\% of tickets relate to
network hardware.  The median incident duration for highest priority
tickets ranges from 114--188 minutes and up to a full day for all
tickets~\cite{turner2012failure}. Other studies confirm these results,
showing median time between connectivity failures over a WAN network
of approximately 3000 seconds with a median time to repair between 2
and 1000 seconds~\cite{ip-backbone-failures} as well as frequent path
routing failures on the Internet~\cite{labovitz-failures}. A recent,
informal report by Kingsbury and Bailis catalogs a host of additional
practitioner reports~\cite{aphyr-post}. Not surprisingly, isolating,
quantifying, and accounting for these network failures is an area of
active research in networking community~\cite{uw-failure-networks}.

These studies indicate that network partitions \textit{do} occur
within and across modern datacenters. We observe that these partitions
must be met with either unavailability at some servers or, as we will
discuss, relaxed semantic guarantees.

\subsection{Latency and Planet Earth}
\label{sec:latency}

Even with fault-free networks, distributed systems face the challenge
of network communication latency, Deutsch's second ``Fallacy.'' In
this section, we quantify round-trip latencies, which are often
large---hundreds of milliseconds in a geo-replicated, multi-datacenter
context.  Fundamentally, the speed at which two servers can
communicate is (according to modern physics) bounded by the speed of
light. In the best case, two servers on opposite sides of the
Earth---communicating via a hypothetical link through the planet's
core---require a minimum 85.1ms round-trip time (RTT; 133.7ms if sent
at surface level). As services are replicated to multiple,
geographically distinct sites, this cost of communication increases.

\definecolor{min-lat-color}{HTML}{B2FF99}
\definecolor{max-lat-color}{HTML}{FF7F7F}

\begin{table}[t!]
\vspace{.5em}

\subfloat[Within \texttt{us-east-b} AZ] {
  \makebox[.2\textwidth]{
 \begin{tabular}{|c|c|c|}\hline
 & \multicolumn{1}{c}{H2} & \multicolumn{1}{c|}{H3}\\\hline
H1  & 0.55   & \colorbox{max-lat-color}{0.56} \\
H2 &  & \colorbox{min-lat-color}{0.50}  \\
\hline
  \end{tabular}}
 }
\subfloat[Across \texttt{us-east} AZs]{
  \makebox[.25\textwidth]{
    \begin{tabular}{|c|c|c|}\hline
 & \multicolumn{1}{c}{C} & \multicolumn{1}{c|}{D}\\\hline
B & \colorbox{min-lat-color}{1.08} & 3.12 \\
C & & \colorbox{max-lat-color}{3.57}  \\
\hline
  \end{tabular}}
}

\subfloat[Cross-region (CA:~California, OR:~Oregon, VA:~Virginia, TO:~Tokyo, IR:~Ireland, SY:~Sydney, SP:~S\~{a}o Paulo, SI:~Singapore)\vspace{-.5em}] {
  \begin{tabular}{|c|c|c|c|c|c|c|c|c|}
\hline
& \multicolumn{1}{c}{OR} & \multicolumn{1}{c}{VA} & \multicolumn{1}{c}{TO} & \multicolumn{1}{c}{IR} & \multicolumn{1}{c}{SY} & \multicolumn{1}{c}{SP} & \multicolumn{1}{c|}{SI} \\\hline
CA & \colorbox{min-lat-color}{22.5}   & 84.5   & 143.7   & 169.8   & 179.1   & 185.9   & 186.9  \\
OR &  & 82.9   & 135.1   & 170.6   & 200.6   & 207.8   & 234.4  \\
VA & &  & 202.4   & 107.9   & 265.6   & 163.4   & 253.5  \\
TO & & &  & 278.3   & 144.2   & 301.4   & 90.6  \\
IR & & & &  & 346.2   & 239.8   & 234.1  \\
SY & & & & &  & 333.6   & 243.1  \\
SP & & & & & &  & \colorbox{max-lat-color}{362.8}  \\
\hline
  \end{tabular}
}

\caption{Mean RTT times on EC2 (min and max highlighted)}\vspace{-1em}
\label{table:rtt}
\end{table}

\begin{figure}[t!]
\includegraphics[width=\columnwidth]{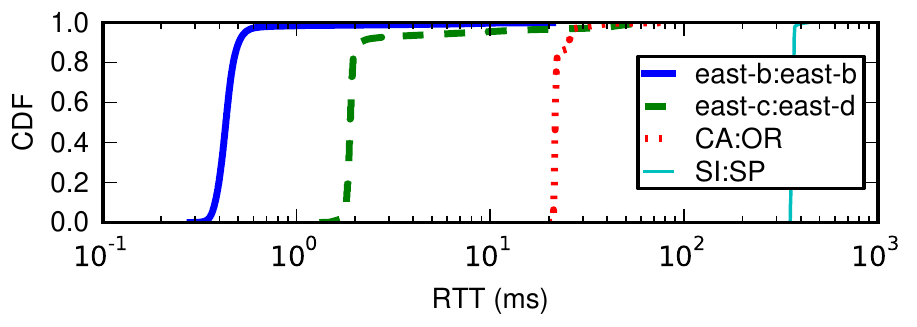}\vspace{-1em}
\caption{CDF of round-trip times for slowest inter- and intra-
  availability zone links compared to cross-region links.}\vspace{-1em}
\label{fig:rtt}
\end{figure}

In real deployments, messages travel slower than the speed of light
due to routing, congestion, and server-side overheads. To illustrate
the difference between intra-datacenter, inter-datacenter, and
inter-planetary networks, we performed a measurement study of network
behavior on Amazon's EC2, a widely used public compute cloud. We
measured one week of ping times (i.e., round-trip times, or RTTs)
between all seven EC2 geographic ``regions,'' across three
``availability zones'' (closely co-located datacenters), and within a
single ``availability zone'' (datacenter), at a granularity of
1s\footnote{Data available at {\scriptsize
    \url{http://bailis.org/blog/communication-costs-in-real-world-networks/},
    \url{https://github.com/pbailis/aws-ping-traces}}}. We summarize
the results of our network measurement study in
Table~\ref{table:rtt}. On average, intra-datacenter communication
(Table 1a) is between 1.82 and 6.38 times faster than across
geographically co-located datacenters (Table 1b) and between 40 and
647 times faster than across geographically distributed datacenters
(Table 1c). The cost of wide-area communication exceeds the speed of
light: for example, while a speed-of-light RTT from S\~{a}o Paulo to
Singapore RTT is 106.7ms, ping packets incur an average 362.8ms RTT
(95th percentile: 649ms). As shown in Figure~\ref{fig:rtt}, the
distribution of latencies varies between links, but the trend is
clear: remote communication has a substantial cost. Quantifying and
minimizing communication delays is also an active area of research in
the networking community~\cite{bobtail}.

\section{ACID in the Wild}
\label{sec:modernacid}

The previous section demonstrated that distributed systems must
address partitions and latency: what does this mean for distributed
databases? Database researchers and designers have long realized that
serializability is not achievable in a highly available
system~\cite{davidson-survey}, meaning that, in environments like
those in Section~\ref{sec:motivation}, database designs face a choice
between availability and strong semantics. However, even in a
single-node database, the coordination penalties associated with
serializability can be severe and are manifested in the form of
decreased concurrency (and, subsequently, performance degradation,
scalability limitations, and, often, aborts due to deadlock or
contention)~\cite{gray-isolation}. Accordingly, to increase
concurrency, database systems offer a range of ACID properties weaker
than serializability: the host of so-called \textit{weak isolation}
models describe varying restrictions on the space of schedules that
are allowable by the system~\cite{adya, ansi-sql, ansicritique}. None
of these weak isolation models guarantees serializability, but, as we
see below, their benefits are often considered to outweigh costs of
possible consistency anomalies that might arise from their use.

To understand the prevalence of weak isolation, we
recently surveyed the default and maximum isolation guarantees provided by
18 databases, often claiming to provide ``ACID'' or ``NewSQL''
functionality~\cite{hat-hotos}. As shown in
Table~\ref{table:existing}, only three out of 18 databases provided
serializability by default, and eight did not provide serializability
as an option at all. This is particularly surprising when we consider
the widespread deployment of many of these non-serializable databases,
like Oracle 11g, which are known to power major businesses and product
functionality. Given that these weak transactional models are frequently
used, our inability to provide serializability in arbitrary HATs
appears non-fatal for practical applications. If application writers
and database vendors have already decided that the benefits of weak
isolation outweigh potential application inconsistencies, then, in a
highly available environment that prohibits serializability, similar
decisions may be tenable.

\begin{table}
\begin{center}
\begin{small}
\begin{tabular}{|l|c|c|}
\hline
Database & Default & Maximum\\\hline
Actian Ingres 10.0/10S & S & S\\
Aerospike & RC & RC\\
Akiban Persistit & SI & SI\\
Clustrix CLX 4100 & RR & RR\\
Greenplum 4.1 & RC & S \\
IBM DB2 10 for z/OS & CS & S\\
IBM Informix 11.50 & Depends & S\\
MySQL 5.6 & RR & S \\
MemSQL 1b & RC & RC\\
MS SQL Server 2012 & RC & S \\
NuoDB & CR & CR\\
Oracle 11g & RC & SI\\
Oracle Berkeley DB & S & S\\
Oracle Berkeley DB JE & RR & S\\
Postgres 9.2.2 & RC & S\\
SAP HANA & RC & SI\\
ScaleDB 1.02 & RC & RC\\
VoltDB & S & S\\
\hline
\multicolumn{3}{|p{7cm}|}{{\begin{small}{RC: read committed, RR: repeatable read, SI: snapshot isolation, S: serializability, CS: cursor stability, CR: consistent read}\end{small}}}\\\hline

\end{tabular}
\caption{Default and maximum isolation levels for ACID and NewSQL
  databases as of January 2013 (from
  \protect\cite{hat-hotos}).}\vspace{-2.5em}
\label{table:existing}
\end{small}
\end{center}
\end{table}

It has been unknown \textit{which} of these guarantees can be provided
with high availability, or are HAT-compliant. Existing algorithms for
providing weak isolation are often designed for a single-node context
and are, to the best of our knowledge, unavailable due to reliance on
concurrency control mechanisms like locking that are not resilient to
partial failure (Section~\ref{sec:eval-existing}). Moreover, we are
not aware of any prior literature that provides guidance as to the
relationship between weak isolation and high availability: prior work
has examined the relationship between serializability and high
availability~\cite{davidson-survey} and weak isolation in
general~\cite{adya, ansicritique, gray-isolation} but not weak
isolation and high availability together.  A primary goal in the
remainder of this paper is to understand which models are
HAT-compliant.

\section{High Availability}
\label{sec:availability}

To understand which guarantees can be provided with high availability,
we must first define what high availability means. In this section, we
will formulate a model that captures a range of availability models,
including high availability, availability with stickiness, and
transactional availability.

Informally, highly available algorithms ensure ``always on'' operation
and, as a side effect, guarantee low latency. If users of a highly
available system are able to contact a (set of) server(s) in a system,
they are guaranteed a response; this means servers will not need to
synchronously communicate with others. If servers are partitioned from one
another, they do not need to stall in order to provide clients a
``safe'' response to operations. This lack of fast-path coordination
also means that a highly available system also provides low
latency~\cite{abadi-pacelc}; in a wide-area setting, clients of a
highly available system need not wait for cross-datacenter
communication. To properly describe whether a \textit{transactional}
system is highly available, we need to describe what servers a client
must contact as well as what kinds of responses a server can provide,
especially given the possibility of aborts.

Traditionally, a system provides {\textbf{high availability}} if every
user that can contact a correct (non-failing) server eventually
receives a response from that server, even in the presence of
arbitrary, indefinitely long network partitions between
servers~\cite{gilbert-cap}.\footnote{Under this definition from the
  distributed systems literature, systems that require a majority of
  servers to be online are not available. Similarly, a system which
  guarantees that servers provide a response with high probability is
  not available. This admittedly stringent requirement matches the
  assumptions made in the CAP Theorem~\cite{gilbert-cap} and
  guarantees low latency~\cite{abadi-pacelc}.} As in a standard
distributed database, designated servers might perform operations for
different data items. A server that can handle an operation for a
given data item is called a \textit{replica} for that
item.\footnote{There is a further distinction between a \textit{fully
    replicated} system, in which all servers are replicas for all data
  items and a \textit{partially replicated} system, in which at least
  one server acts as a replica for a proper subset of all data
  items. For generality, and, given the prevalence of these
  ``sharded'' or ``partitioned'' systems~\cite{ bigtable, pnuts,
    spanner, dynamo, hstore}, we consider partial replication here.}

\subsection{Sticky Availability}
\label{sec:sticky}

In addition to high availability, which allows operations on any
replica, distributed algorithms often assume a model in which clients
always contact the same logical replica(s) across subsequent
operations, whereby each of the client's prior operations (but not
necessarily other clients' operations) are reflected in the database
state that they observe. As we will discuss in Section~\ref{sec:hats},
clients can ensure continuity between operations (e.g., reading their
prior updates to a data item) by maintaining affinity or
``stickiness'' with a server or set of servers~\cite{vogels-defs}. In
a fully replicated system, where all servers are replicas for all data
items, stickiness is simple: a client can maintain stickiness by
contacting the same server for each of its requests. However, to stay
``sticky'' in a partially-replicated system, where servers are
replicas for subsets of the set of data items (which we consider in
this paper), a client must maintain stickiness with a single
\textit{logical} copy of the database, which may consist of multiple
physical servers. We say that a system provides \textbf{sticky
  availability} if, whenever a client's transactions is executed
against a copy of database state that reflects all of the client's
prior operations, it eventually receives a response, even in the
presence of indefinitely long partitions (where ``reflects'' is
dependent on semantics). A client may choose to become sticky
available by acting as a server itself; for example, a client might
cache its reads and writes~\cite{bolton, sessionguarantees,
  swift}. Any guarantee achievable in a highly available system is
achievable in a sticky high availability system but not vice-versa.

\subsection{Transactional Availability}

Until now, we have considered single-object, single-operation
availability. This is standard in the distributed systems literature
(e.g., distributed register models such as linearizability all concern
single objects~\cite{herlihy-art}), yet the database literature
largely focuses on transactions: groups of multiple operations over
multiple objects. Accordingly, by itself, traditional definitions of
high availability are insufficient to describe availability guarantees
for transactions. Additionally, given the choice of \textit{commit}
and \textit{abort} responses---which signal transaction success or
failure to a client---we must take care in defining transactional
availability.

We say that a transaction has \textbf{replica availability} if it can
contact at least one replica for every item it attempts to access;
this may result in ``lower availability'' than a non-transactional
availability requirement (e.g., single-item
availability). Additionally, given the possibility of system-initiated
aborts, we need to ensure useful forward progress: a system can
trivially guarantee clients a response by always aborting all
transactions. However, this is an unsatisfactory system because
nothing good (transaction commit) ever happens; we should require a
\textit{liveness} property~\cite{transaction-liveness}.

A system cannot guarantee that every transaction will
commit---transactions may choose to abort themselves---but we need to
make sure that the system will not indefinitely abort transactions on
its own volition. We call a transaction abort due to a transaction's
own choosing (e.g., as an operation of the transaction itself or due
to a would-be violation of a declared integrity constraint) an
\textit{internal abort} and an abort due to system implementation or
operation an \textit{external abort}. We say that a system provides
\textbf{transactional availability} if, given replica availability for
every data item in a transaction, the transaction eventually commits
(possibly after multiple client retries) or internally
aborts~\cite{hat-hotos}. A system provides \textbf{sticky
  transactional availability} if, given sticky availability, a
transaction eventually commits or internally aborts.

\section{Highly Available Transactions}
\label{sec:hats}

HAT systems provide transactions with transactional availability or
sticky transactional availability. They offer latency and availability
benefits over traditional distributed databases, yet they cannot
achieve all possible semantics. In this section, we describe ACID,
distributed replica consistency, and session consistency levels which
can be achieved with high availability (Read Committed isolation,
variants of Repeatable Read, atomic reads, and many session
guarantees), those with sticky availability (read your writes, PRAM
and causal consistency). We also discuss properties that cannot be
provided in a HAT system (those preventing Lost Update and Write Skew
or guaranteeing recency).  We present a full summary of these results
in Section~\ref{sec:hat-summary}.

As Brewer states, ``systems and database communities are separate but
overlapping (with distinct vocabulary)''~\cite{brewer-slides}. With
this challenge in mind, we build on existing properties and
definitions from the database and distributed systems literature,
providing a brief, informal explanation and example for each
guarantee. The database isolation guarantees require particular care,
since different DBMSs often use the same terminology for different
mechanisms and may provide additional guarantees in addition to our
implementation-agnostic definitions.  We draw largely on Adya's
dissertation~\cite{adya} and somewhat on its predecessor work: the
ANSI SQL specification~\cite{ansi-sql} and Berenson et al.'s
subsequent critique~\cite{ansicritique}.

For brevity, we provide an informal presentation of each guarantee
here (accompanied by appropriate references) but give a full set of
formal definitions in
\iftechreport
 Appendix A.
\else
 our extended Technical Report~\cite{hat-tr}.
\fi
In our examples, we exclusively consider read and write operations,
denoting a write of value $v$ to data item $d$ as $w_d(v)$ and a read
from data item $d$ returning $v$ as $r_d(v)$. We assume that all data
items have the null value, $\bot$, at database initialization, and,
unless otherwise specified, all transactions in the examples commit.

\subsection{Achievable HAT Semantics}

To begin, we present well-known semantics that can be achieved in HAT
systems. In this section, our primary goal is feasibility, not
performance. As a result, we offer proof-of-concept highly available
algorithms that are not necessarily optimal or even efficient: the
challenge is to prove the existence of algorithms that provide high
availability. However, we briefly study a subset of their performance
implications in Section~\ref{sec:evaluation}.

\subsubsection{ACID Isolation Guarantees}
\label{sec:isolation}

To begin, Adya captures \textbf{Read Uncommitted} isolation as
\textit{PL-1}. In this model, writes to each object are totally
ordered, corresponding to the order in which they are installed in the
database. In a distributed database, different replicas may receive
writes to their local copies of data at different times but should
handle concurrent updates (i.e., overwrites) in accordance with the
total order for each item. \textit{PL-1} requires that writes to
different objects be ordered consistently across transactions,
prohibiting Adya's phenomenon $G0$ (also called ``Dirty
Writes''~\cite{ansicritique}). If we build a graph of transactions
with edges from one transaction to another and, when the former
overwrites the latter's write to the same object, then, under Read
Uncommitted, the graph should not contain cycles~\cite{adya}. Consider
the following example:
\begin{align*}
\small\vspace{-1em}
T_1 &: w_x(1)~w_y(1)
\\T_2 &: w_x(2)~w_y(2)
\end{align*}
In this example, under Read Uncommitted, it is impossible for the
database to order $T_1$'s $w_x(1)$ before $T_2$'s $w_x(2)$ but order
$T_2$'s $w_y(2)$ before $T_1$'s $w_y(1)$. Read Uncommitted is easily
achieved by marking each of a transaction's writes with the same
timestamp (unique across transactions; e.g., combining a client's ID
with a sequence number) and applying a ``last writer wins'' conflict
reconciliation policy at each replica. Later properties will
strengthen Read Uncommitted.

\textbf{Read Committed} isolation is particularly important in
practice as it is the default isolation level of many DBMSs
(Section~\ref{sec:modernacid}). Centralized implementations differ,
with some based on long-duration exclusive locks and short-duration
read locks~\cite{gray-isolation} and others based on multiple
versions. These implementations often provide recency and monotonicity
properties beyond what is implied by the name ``Read Committed'' and
what is captured by the implementation-agnostic definition: under Read
Committed, transactions should not access uncommitted or intermediate
versions of data items. This prohibits both ``Dirty Writes'', as
above, and also ``Dirty Reads'' phenomena.  This isolation is Adya's
\textit{PL-2} and is formalized by prohibiting Adya's
\textit{G1\{a-c\}} (or ANSI's $P1$, or ``broad'' $P1$ [2.2] from
Berenson et al.). For instance, in the example below, $T_3$ should
never see $a=1$, and, if $T_2$ aborts, $T_3$ should not read $a=3$:
\begin{align*}
\small\vspace{-1em}
T_1 &: w_x(1)~w_x(2)
\\T_2 &: w_x(3)\\
T_3 &: r_x(a)\vspace{-1em}
\end{align*}
It is fairly easy for a HAT system to prevent ``Dirty Reads'': if each
client never writes uncommitted data to shared copies of data, then
transactions will never read each others' dirty data. As a simple
solution, clients can buffer their writes until they commit, or,
alternatively, can send them to servers, who will not deliver their
value to other readers until notified that the writes have been
committed. Unlike a lock-based implementation, this implementation
does not provide recency or monotonicity guarantees but it satisfies
the implementation-agnostic definition.

Several different properties have been labeled \textbf{Repeatable
  Read} isolation. As we will show in
Section~\ref{sec:unachievable-acid}, some of these are not achievable
in a HAT system. However, the ANSI standardized
implementation-agnostic definition~\cite{ansi-sql} \textit{is}
achievable and directly captures the spirit of the term: if a
transaction reads the same data more than once, it sees the same value each
time (preventing ``Fuzzy Read,'' or $P2$). In this paper, to
disambiguate between other definitions of ``Repeatable Read,'' we will
call this property ``cut isolation,'' since each transaction reads
from a non-changing cut, or snapshot, over the data items. If this
property holds over reads from discrete data items, we call it
\textbf{Item Cut Isolation}, and, if we also expect a cut over
predicate-based reads (e.g., \texttt{SELECT WHERE}; preventing
Phantoms~\cite{gray-isolation}, or Berenson et al.'s $P3/A3$), we have
the stronger property of \textbf{Predicate Cut-Isolation}. In the
example below, under both levels of cut isolation, $T_3$ must read
$a=1$:
\begin{align*}
\small
T_1 &: w_x(1)
\\T_2 &: w_x(2)
\\T_3 &: r_x(1)~r_x(a)
\end{align*}
It is possible to satisfy Item Cut Isolation with high availability by
having transactions store a copy of any read data at the client such
that the values that they read for each item never changes unless they
overwrite it themselves. These stored values can be discarded at the
end of each transaction and can alternatively be accomplished on
(sticky) replicas via multi-versioning. Predicate Cut Isolation is
also achievable in HAT systems via similar caching middleware or
multi-versioning that track entire logical ranges of predicates in
addition to item based reads.

\subsubsection{ACID Atomicity Guarantees}
\label{sec:ta}

Atomicity, informally guaranteeing that either all or none of
transactions' effects should succeed, is core to ACID
guarantees. Although, at least by the ACID acronym, atomicity is not
an ``isolation'' property, atomicity properties also restrict the
updates visible to other transactions. Accordingly, here, we consider
the \textit{isolation} effects of atomicity, which we call
\textbf{Monotonic Atomic View (MAV)} isolation.  Under MAV, once some of the
effects of a transaction $T_i$ are observed by another transaction
$T_j$, thereafter, all effects of $T_i$ are observed by $T_j$. That
is, if a transaction $T_j$ reads a version of an object that
transaction $T_i$ wrote, then a later read by $T_j$ cannot return a
value whose later version is installed by $T_i$. Together with item
cut isolation, MAV prevents Read Skew anomalies (Berenson et al.'s A5A)
and is useful in several contexts such as maintaining foreign key
constraints, consistent global secondary indexing, and maintenance of
derived data. In the example below, under MAV, because $T_2$ has read
$T_1$'s write to $y$, $T_2$ must observe $b=c=1$ (or later versions
for each key):
\begin{align*}
\small
T_1 &: w_x(1)~w_y(1)~w_z(1)
\\T_2 &: r_x(a)~r_y(1)~r_x(b)~r_z(c)~\\[-1.5em]
\end{align*}
$T_2$ can also observe $a=\bot$, $a=1$, or a later version of $x$. In
the hierarchy of existing isolation properties, we place MAV below
Adya's \textit{PL-2L} (as it does not necessarily enforce transitive
read-write dependencies) but above Read Committed ($PL-2$). Notably,
MAV requires disallows reading intermediate writes (Adya's $G1b$):
observing all effects of a transaction implicitly requires observing
the final (committed) effects of the transaction as well.

Perplexingly, discussions of MAV are absent from existing treatments of
weak isolation. This is perhaps again due to the single-node context
in which prior work was developed: on a single server (or a fully
replicated database), MAV is achievable via lightweight locking and/or
local concurrency control over data items~\cite{gstore,
  kemme-thesis}. In contrast, in a distributed environment, MAV over
arbitrary groups of non-co-located items is considerably more difficult
to achieve with high availability.


As a straw man, replicas can store all versions ever written to each
data item. Replicas can gossip information about versions they have
observed and construct a lower bound on the versions that can be found
on every replica (which can be represented by either a list of
versions, or, more realistically, a vector clock). At the start of
each transaction, clients can choose a \textit{read timestamp} that is
lower than or equal to the this global lower bound, and, during
transaction execution, replicas return the latest version of each item
that is not greater than the client's chosen timestamp. If this lower
bound is advanced along transactional boundaries, clients will observe
MAV. This algorithm has several variants in the
literature~\cite{readonly, swift}, and older versions can be
asynchronously garbage collected.

We have developed a more efficient MAV algorithm, which we sketch here
and provide greater detail in \iftechreport Appendix B.  \else our
extended Technical Report~\cite{hat-tr}.  \fi We begin with our Read
Committed algorithm, but replicas wait to reveal new writes to readers
until all of the replicas for the final writes in the transaction have
received their respective writes (are \textit{pending
  stable}). Clients include additional metadata with each write: a
single timestamp for all writes in the transaction (e.g., as in Read
Uncommitted) and a list of items written to in the transaction. When a
client reads, the return value's timestamp and list of items form a
lower bound on the versions that the client should read for the other
items. When a client reads, it attaches a timestamp to its request
representing the current lower bound for that item. Replicas use this
timestamp to respond with either a write matching the timestamp or a
pending stable write with a higher timestamp. Servers keep two sets of
writes for each data item: the write with the highest timestamp that
is pending stable and a set of writes that are not yet pending
stable. This is entirely master-less and operations never block due to
replica coordination.

\subsubsection{Session Guarantees}

A useful class of safety guarantees refer to real-time or
client-centric ordering within a \textit{session}, ``an abstraction
for the sequence of...operations performed during the execution of an
application''~\cite{sessionguarantees}. These ``session guarantees''
have been explored in the distributed systems
literature~\cite{sessionguarantees,vogels-defs} and sometimes in the
database literature~\cite{daudjee-session}. For us, a session
describes a context that should persist between transactions: for
example, on a social networking site, all of a user's transactions
submitted between ``log in'' and ``log out'' operations might form a
session.

Several session guarantees can be made with high availability:

\vspace{.5em}\noindent\textbf{{Monotonic reads}} requires that, within
a session, subsequent reads to a given object ``never return any
previous values''; reads from each item progress according to a total
order (e.g., the order from Read Uncommitted).

\vspace{.5em}\noindent\textbf{{Monotonic writes}} requires that each
session's writes become visible in the order they were submitted. Any
order on transactions (as in Read Uncommitted isolation) should also
be consistent with any precedence that a global observer would see.

\vspace{.5em}\noindent\textbf{{Writes Follow Reads}} requires that, if
a session observes an effect of transaction $T_1$ and subsequently
commits transaction $T_2$, then another session can only observe
effects of $T_2$ if it can also observe $T_1$'s effects (or later
values that supersede $T_1$'s); this corresponds to Lamport's
``happens-before'' relation~\cite{lamportclocks}.  Any order on
transactions should respect this transitive order.\vspace{.5em}

The above guarantees can be achieved by forcing servers to wait to
reveal new writes (say, by buffering them in separate local storage)
until each write's respective dependencies are visible on all
replicas. This mechanism effectively ensures that all clients read
from a globally agreed upon lower bound on the versions written. This
is highly available because a client will never block due to inability
to find a server with a sufficiently up-to-date version of a data
item. However, it does not imply that transactions will read their own
writes or, in the presence of partitions, make forward progress
through the version history. The problem is that under non-sticky
availability, a system must handle the possibility that, under a
partition, an unfortunate client will be forced to issue its next
requests against a partitioned, out-of-date server.

A solution to this conundrum is to forgo high availability and settle
for sticky availability. Sticky availability permits three additional
guarantees, which we first define and then prove are unachievable in a
generic highly available system:

\vspace{.5em}\noindent\textbf{{Read your writes}} requires that
whenever a client reads a given data item after updating it, the read
returns the updated value (or a value that overwrote the previously
written value).

\vspace{.5em}\noindent\textbf{{PRAM}} (Pipelined Random Access Memory)
provides the illusion of serializing each of the operations (both
reads and writes) within each session and is the combination of
monotonic reads, monotonic writes, and read your
writes~\cite{herlihy-art}.

\vspace{.5em}\noindent\textbf{{Causal
    consistency}}~\cite{causalmemory} is the combination of all of the
session guarantees~\cite{sessiontocausal} (alternatively, PRAM with
writes-follow-reads) and is also referred to by Adya as \textit{PL-2L}
isolation~\cite{adya}).\vspace{.5em}

Read your writes is not achievable in a highly available
system. Consider a client that executes the following two transactions:
\begin{align*}
\small
T_1 &: w_x(1)
\\T_2 &: r_x(a)
\end{align*}
If the client executes $T_1$ against a server that is partitioned from
the rest of the other servers, then, for transactional availability,
the server must allow $T_1$ to commit. If the same client subsequently
executes $T_2$ against the same (partitioned) server in the same
session, then it will be able to read its writes. However, if the
network topology changes and the client can only execute $T_2$ on a
different replica that is partitioned from the replica that executed
$T_1$, then the system will have to either stall indefinitely to allow
the client to read her writes (violating transactional availability)
or will have to sacrifice read your writes guarantees. However, if the
client remains sticky with the server that executed $T_1$, then we can
disallow this scenario. Accordingly, read your writes, and, by proxy,
causal consistency and PRAM require stickiness. Read your writes is
provided by default in a sticky system. Causality and PRAM guarantees
can be accomplished with well-known variants~\cite{causalmemory,
  bolton, eiger, sessionguarantees, swift} of the prior session
guarantee algorithms we presented earlier: only reveal new writes to
clients when their (respective, model-specific) dependencies have been revealed.

\subsubsection{Additional HAT Guarantees}

In this section, we briefly discuss two additional kinds of guarantees
that are achievable in HAT systems.

\vspace{0.5em}
\noindent{\textbf{Consistency}} A HAT system can make limited
application-level consistency guarantees. It can often execute
commutative and logically monotonic~\cite{calm} operations without the
risk of invalidating application-level integrity constraints and can
maintain limited criteria like foreign key constraints (via MAV). We do
not describe the entire space of application-level consistency
properties that are achievable (see Section~\ref{sec:relatedwork}) but
we specifically evaluate TPC-C transaction semantics with HAT
guarantees in Section~\ref{sec:evaluation}.

\vspace{.5em}\noindent{\textbf{Convergence} Under arbitrary (but not
  infinite delays), HAT systems can ensure convergence, or
  \textit{eventual consistency}: in the absence of new mutations to a
  data item, all servers should eventually agree on the value for each
  item~\cite{cac, vogels-defs}. This is typically accomplished by any
  number of anti-entropy protocols, which periodically update
  neighboring servers with the latest value for each data
  item~\cite{antientropy}. Establishing a final convergent value is
  related to determining a total order on transaction updates to each
  item, as in Read Uncommitted.

\subsection{Unachievable HAT Semantics}
\label{sec:unachievable-hat}

While there are infinitely many HAT models
(Section~\ref{sec:relatedwork}), at this point, we have largely
exhausted the range of achievable, previously defined (and useful)
semantics that are available to HAT systems. Before summarizing our
possibility results, we will present impossibility results for HATs,
also defined in terms of previously identified isolation and
consistency anomalies. Most notably, it is impossible to
prevent Lost Update or Write Skew in a HAT system.

\subsubsection{Unachievable ACID Isolation}
\label{sec:unachievable-acid}

In this section, we demonstrate that preventing Lost Update and Write
Skew---and therefore providing Snapshot Isolation, Repeatable Read,
and one-copy serializability---inherently requires foregoing high
availability guarantees.

Berenson et al. define \textit{Lost Update} as when one
transaction $T1$ reads a given data item, a second transaction $T2$
updates the same data item, then $T1$ modifies the data item based on
its original read of the data item, ``missing'' or ``losing'' $T2$'s
newer update. Consider a database containing only the following
transactions:
\begin{align*}
\small\vspace{-1em}
T_1 &: r_x(a)~w_x(a+2)
\\T_2 &: w_x(2)\vspace{-1em}
\end{align*}
If $T_1$ reads $a=1$ but $T_2$'s write to $x$ precedes $T_1$'s write
operation, then the database will end up with $a=3$, a state that
could not have resulted in a serial execution due to $T_2$'s
``Lost Update.''

It is impossible to prevent Lost Update in a highly available
environment. Consider two clients who submit the following $T_1$ and
$T_2$ on opposite sides of a network partition:
\begin{align*}
\small\vspace{-1em}
T_1 &: r_x(100)~w_x(100+20=120)
\\T_2 &: r_x(100)~w_x(100+30=130)\vspace{-1em}
\end{align*}
Regardless of whether $x=120$ or $x=130$ is chosen by a replica, the
database state could not have arisen from a serial execution of $T_1$ and
$T_2$.\footnote{In this example, we assume that, as is standard in
  modern databases, replicas accept values as they are written (i.e.,
  register semantics). This particular example could be made
  serializable via the use of commutative updates
  (Section~\ref{sec:evaluation}) but the problem persists in the
  general case.}  To prevent this, either $T_1$ or
$T_2$ should not have committed. Each client's respective server might
try to detect that another write occurred, but this requires knowing
the version of the latest write to $x$. In our example, this reduces
to a requirement for linearizability, which is, via Gilbert and
Lynch's proof of the CAP Theorem, provably at odds with high
availability~\cite{gilbert-cap}.

\textbf{Write Skew} is a generalization of Lost Update to multiple
keys. It occurs when one transaction $T1$ reads a given data item $x$,
a second transaction $T2$ reads a different data item $y$, then $T1$
writes to $y$ and commits and $T2$ writes to $x$ and commits. As an
example of Write Skew, consider the following two transactions:
\begin{align*}
\small
T_1 &: r_y(0)~w_x(1)
\\T_2 &: r_x(0)~w_y(1)
\end{align*}
As Berenson et al. describe, if there was an integrity constraint
between $x$ and $y$ such that only one of $x$ or $y$ should have value
$1$ at any given time, then this write skew would violate the constraint (which is preserved in serializable executions). Write skew is a somewhat
esoteric anomaly---for example, it does not appear in
TPC-C~\cite{snapshot-serializable}---but, as a generalization of Lost
Update, it is also unavailable to HAT systems.

Consistent Read, Snapshot Isolation (including Parallel Snapshot
Isolation~\cite{walter}), and Cursor Stability guarantees are all
unavailable because they require preventing Lost Update phenomena.
Repeatable Read (defined by Gray~\cite{gray-isolation}, Berenson et
al.~\cite{ansicritique}, and Adya~\cite{adya}) and One-Copy
Serializability~\cite{1sr} need to prevent both Lost Update and Write
Skew. Their prevention requirements mean that these guarantees are
inherently unachievable in a HAT system.

\subsubsection{Unachievable Recency Guarantees}

Distributed data storage systems often make various recency guarantees
on reads of data items.  Unfortunately, an indefinitely long partition
can force an available system to violate any recency bound, so recency
bounds are not enforceable by HAT systems~\cite{gilbert-cap}. One of
the most famous of these guarantees is
linearizability~\cite{herlihy-art}, which states that reads will
return the last completed write to a data item, and there are several
other (weaker) variants such as safe and regular register
semantics. When applied to transactional semantics, the combination of
one-copy serializability and linearizability is called \textit{strong
  (or strict) one-copy serializability}~\cite{adya} (e.g.,
Spanner~\cite{spanner}). It is also common, particularly in systems
that allow reading from masters and slaves, to provide a guarantee
such as ``read a version that is no more than five seconds out of
date'' or similar. None of these guarantees are HAT-compliant.

\subsubsection{Durability}

A client requiring that its transactions' effects survive $F$ server
faults requires that the client be able to contact at least $F+1$
non-failing replicas before committing. This affects
availability and, according to the Gilbert and Lynch definition we
have adopted, $F>1$ fault tolerance is not achievable with high
availability.

\subsection{Summary}
\label{sec:hat-summary}

As we summarize in Table~\ref{table:hatcompared}, a wide range of
isolation levels are achievable in HAT systems. With sticky
availability, a system can achieve read your writes guarantees and
PRAM and causal consistency. However, many other prominent semantics,
such as Snapshot Isolation, One-Copy Serializability, and Strong
Serializability cannot be achieved due to the inability to prevent
Lost Update and Write Skew phenomena.

We illustrate the hierarchy of available, sticky available, and
unavailable consistency models we have discussed in
Figure~\ref{fig:hatcompared}. Many models are simultaneously
achievable, but we find several particularly compelling. If we combine
all HAT and sticky guarantees, we have transactional, causally
consistent snapshot reads (i.e., Causal Transactional Predicate Cut
Isolation). If we combine MAV and P-CI, we have transactional snapshot
reads. We can achieve RC, MR, and RYW by simply sticking clients to
servers. We can also combine unavailable models---for example, an
unavailable system might provide PRAM and One-Copy
Serializability~\cite{daudjee-session}.

To the best of our knowledge, this is the first unification of
transactional isolation, distributed consistency, and session
guarantee models. Interestingly, strong one-copy serializability
entails all other models, while considering the (large) power set of
all compatible models (e.g., the diagram depicts 144 possible HAT
combinations) hints at the vast expanse of consistency models found in
the literature. This taxonomy is not exhaustive
(Section~\ref{sec:conclusion}), but we believe it lends substantial
clarity to the relationships between a large subset of the prominent
ACID and distributed consistency models. Additional read/write
transaction semantics that we have omitted should be classifiable
based on the available primitives and HAT-incompatible anomaly
prevention we have already discussed.

In light the of current practice of deploying weak isolation levels
(Section~\ref{sec:modernacid}), it is perhaps surprising that so many
weak isolation levels are achievable as HATs. Indeed, isolation levels
such as Read Committed expose and are defined in terms of end-user
anomalies that could not arise during serializable execution. However,
the prevalence of these models suggests that, in many cases,
applications can tolerate these their associated anomalies. Given our
HAT-compliance results, this in turn hints that--despite
idiosyncrasies relating to concurrent updates and data recency--highly
available database systems can provide sufficiently strong semantics
for many applications. Indeed, HAT databases may expose more anomalies
than a single-site database operating under weak isolation
(particularly during network partitions). However, for a fixed
isolation level (which, in practice, can vary across databases and may
differ from implementation-agnostic definitions in the literature),
users of single-site database are subject to the same (worst-case)
application-level anomalies as a HAT implementation. The necessary
(indefinite) visibility penalties (i.e., the right side of
Figure~\ref{fig:hatcompared}) and lack of support for preventing
concurrent updates (via the upper left half of
Figure~\ref{fig:hatcompared}) mean HATs are \textit{not} well-suited
for all applications (see Section~\ref{sec:evaluation}): these
limitations are fundamental. However, common practices such as ad-hoc,
user-level compensation and per-statement isolation ``upgrades''
(e.g., \texttt{SELECT FOR UPDATE} under weak isolation)---commonly used
to augment weak isolation---are also applicable in HAT systems
(although they may in turn compromise availability).

 \newcommand{\lostupdate}{$^\dagger$}
 \newcommand{\rwskew}{$^\ddagger$}
 \newcommand{\linearizable}{$^\oplus$}

\begin{table}[t!]
\begin{tabular}{| c | p{6cm} | }\hline
HA & Read Uncommitted (RU), Read Committed (RC), Monotonic Atomic View
(MAV), Item Cut Isolation (I-CI), Predicate Cut Isolation (P-CI),
Writes Follow Reads (WFR), Monotonic Reads (MR), Monotonic Writes
(MW)\\\hline Sticky & Read Your Writes (RYW), PRAM, Causal\\\hline
Unavailable & Cursor Stability (CS)\lostupdate, Snapshot Isolation
(SI)\lostupdate, Repeatable Read (RR)\lostupdate\rwskew, One-Copy
Serializability (1SR)\lostupdate\rwskew, Recency\linearizable,
Safe\linearizable, Regular\linearizable, Linearizability\linearizable,
Strong 1SR\lostupdate\rwskew\linearizable \\\hline
\end{tabular}
\caption{Summary of highly available, sticky available, and
  unavailable models considered in this paper. Unavailable models are
  labeled by cause of unavailability: preventing lost
  update\lostupdate, preventing write skew\rwskew, and requiring
  recency guarantees\linearizable.}
\label{table:hatcompared}
\end{table}

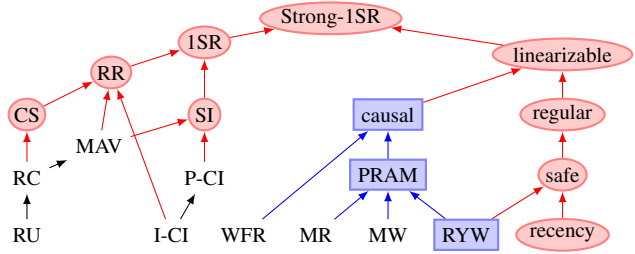
\begin{figure}[t!]
\centering
\begin{tikzpicture}[scale=0.8]
  \tikzstyle{sticky}=[rectangle,draw=blue!50,fill=blue!20,thick]
  \tikzstyle{noha}=[ellipse,draw=red!50,fill=red!20,thick, inner sep=0pt,minimum size=12pt]

  \tikzstyle{every node}=[font=\small]

 \node[draw=none,fill=none] (ici) at (1.2, 0) {I-CI};
 \node[draw=none,fill=none] (pci) at (1.75, .95) {P-CI};
 \node[draw=none,fill=none] (rc) at (-1.2, .95) {RC};
 \node[draw=none,fill=none] (ru) at (-1.2, 0) {RU};

 \node[draw=none,fill=none] (ra) at (0, 1.475) {MAV};

 \node[draw=none,fill=none] (mr) at (3.6, 0) {MR};
 \node[draw=none,fill=none] (mw) at (4.8, 0) {MW};
 \node[draw=none,fill=none] (wfr) at (2.4,0) {WFR};
 \node at (6.1,0) [sticky] (ryw) {RYW};

 \node[noha](recency) at (7.7, 0) {recency};
 \node[noha](safe) at (7.7, 1) {safe};
 \node[noha](regular) at (7.7, 2) {regular};
 \node[noha](linearizable) at (7.7, 3) {linearizable};
 \node at (4.8, 2) [sticky] (causal) {causal};
 \node at (4.8, 1) [sticky] (pram) {PRAM};
 \node[noha] (cs) at (-1.2, 2) {CS};
 \node[noha] (rr) at (0.2, 2.7) {RR};
 \node[noha] (si) at (1.75, 2) {SI};
 \node[noha] (1sr) at (1.75, 3.2) {1SR};
 \node[noha] (ssr) at (3.85, 3.6) {Strong-1SR};

 \draw [->, red] (recency) -- (safe);
 \draw [->, red] (safe) -- (regular);
 \draw [->, red] (regular) -- (linearizable);
 \draw [->, red] (linearizable) -- (ssr);
 \draw [->, red] (1sr) -- (ssr);
 
 \draw [->] (ru) -- (rc);
 \draw [->] (rc) -- (ra);
 \draw [->] (ici) -- (pci);

 \draw [->, blue] (mr) -- (pram);
 \draw [->, blue] (mw) -- (pram);
 \draw [->, blue] (wfr) -- (causal);
 \draw [->, blue] (ryw) -- (pram);
 \draw [->, blue] (pram) -- (causal);


 \draw [->, red] (rc) -- (cs);
 \draw [->, red] (cs) -- (rr);
 \draw [->, red] (pci) -- (si);
 \draw [->, red] (ici) -- (rr);
 \draw [->, red] (rr) -- (1sr);
 \draw [->, red] (si) -- (1sr);
 \draw [->, red] (ra) -- (si);
 \draw [->, red] (ra) -- (rr);
 \draw [->, red] (causal) -- (linearizable);
 \draw [->, red] (ryw) -- (safe);

\end{tikzpicture}
\label{fig:hat-order}
\caption{Partial ordering of HAT, sticky available (in boxes, blue), and
  unavailable models (circled, red) from
  Table~\protect\ref{table:hatcompared}. Directed edges represent
  ordering by model strength. Incomparable models can be
  simultaneously achieved, and the availability of a combination of
  models has the availability of the least available individual
  model.}\vspace{-1em}
\label{fig:hatcompared}
\end{figure}

\section{HAT Implications}
\label{sec:evaluation}

With an understanding of which semantics are HAT-compliant, in this
section, we analyze the implications of these results for existing
systems and briefly study HAT systems on public cloud
infrastructure. Specifically:
\begin{myenumerate}
\item We revisit traditional database concurrency control with a focus
  on coordination costs and on high availability.
\item We examine the properties required by an OLTP application based
  on the TPC-C benchmark.
\item We perform a brief experimental evaluation of HAT versus non-HAT
  properties on public cloud infrastructure.
\end{myenumerate}

\subsection{HA and Existing Algorithms}
\label{sec:eval-existing}

While we have shown that many database isolation levels are achievable
as HATs, many traditional concurrency control mechanisms do not
provide high availability---even for HAT-compliant isolation
levels. Existing mechanisms often presume (or are adapted from)
single-server non-partitioned deployments or otherwise focus on
serializability as a primary use case. In this section, we briefly
discuss design decisions and algorithmic details that preclude high
availability.

\vspace{.5em}\noindent\textbf{Serializability} To establish a serial
order on transactions, algorithms for achieving serializability of
general-purpose read-write transactions in a distributed
setting~\cite{bernstein-book, davidson-survey} require at least one
RTT before committing. As an example, traditional two-phase locking
for a transaction of length $T$ may require $T$ \texttt{lock}
operations and will require at least one \texttt{lock} and one
\texttt{unlock} operation.  In a distributed environment, each of
these lock operations requires coordination, either with other
database servers or with a lock service. If this coordination
mechanism is unavailable, transactions cannot safely
commit. Similarly, optimistic concurrency control requires
coordinating via a validation step, while deterministic transaction
scheduling~\cite{deterministic-scheduling} requires contacting a
scheduler. Serializability under multi-version concurrency control
requires checking for update conflicts. All told, the reliance on a
globally agreed total order necessitates a minimum of one round-trip
to a designated master or coordination service for each of these
classic algorithms.  As we saw in Section~\ref{sec:motivation}, is
will be determined by the deployment environment; we will further
demonstrate this in Section~\ref{sec:prototype}.

\vspace{.5em}\noindent\textbf{Non-serializability} Most existing
distributed implementations of weak isolation are not highly
available. Lock-based mechanisms such as those in Gray's original
proposal~\cite{gray-isolation} do not degrade gracefully in the
presence of partial failures. (Note, however, that lock-based
protocols \textit{do} offer the benefit of recency guarantees.) While
multi-versioned storage systems allow for a variety of transactional
guarantees, few offer traditional weak isolation (e.g.,
non-``tentative update'' schemes) in this context.  Chan and Gray's
read-only transactions have item-cut isolation with causal consistency
and MAV (session \textit{PL-2L}~\cite{adya}) but are unavailable in
the presence of coordinator failure and assume serializable update
transactions~\cite{readonly}; this is similar to read-only and
write-only transactions more recently proposed by Eiger~\cite{eiger}.
Brantner's S3 database~\cite{kraska-s3} and
Bayou~\cite{sessionguarantees} can all provide variants of session
\textit{PL-2L} with high availability, but none provide this HAT
functionality without substantial modification. Accordingly, it is
possible to implement many guarantees weaker than
serializability---including HAT semantics---and still not achieve high
availability. We view high availability as a core design consideration
in future concurrency control designs.

\subsection{Application Requirements}

Thus far, we have largely ignored the question of when HAT semantics
are useful (or otherwise are too weak). As we showed in
Section~\ref{sec:hats}, the main cost of high availability and low
latency comes in the inability to prevent Lost Update, Write Skew, and
provide recency bounds. To better understand the impact of
HAT-compliance in an application context, we consider a concrete
application: the TPC-C benchmark. In brief, we find that four of five
transactions can be executed via HATs, while the fifth requires
unavailability.

TPC-C consists of five transactions, capturing the operation of a
wholesale warehouse, including sales, payments, and deliveries. Two
transactions---\textit{Order-Status} and \textit{Stock-Level}---are
read-only and can be executed safely with HATs. Clients may read stale
data, but this does not violate TPC-C requirements and clients will
read their writes if they are sticky-available. Another transaction
type, \textit{Payment}, updates running balances for warehouses,
districts, and customer records and provides an audit trail. The
transaction is monotonic---increment- and append-only---so all balance
increase operations commute, and MAV allows the maintenance of
foreign-key integrity constraints (e.g., via \texttt{UPDATE/DELETE
  CASCADE}).

\vspace{.5em}\noindent\textit{New-Order and Delivery.} While three out of
five transactions are easily achievable with HATs, the remaining two
transactions are not as simple. The New-Order transaction places an
order for a variable quantity of data items, updating warehouse stock
as needed. It selects a sales district, assigns the order an ID
number, adjusts the remaining warehouse stock, and writes a
placeholder entry for the pending order. The Delivery transaction
represents the fulfillment of a New-Order: it deletes the order from
the pending list, updates the customer's balance, updates the order's
carrier ID and delivery time, and updates the customer balance.

\vspace{.5em}\noindent\textit{IDs and decrements.} The New-Order
transaction presents two challenges: ID assignment and stock
maintenance. First, each New-Order transaction requires a unique ID
number for the order. We can create a unique number by, say,
concatenating the client ID and a timestamp. However, the TPC-C
specification requires order numbers to be \textit{sequentially}
assigned within a district, which requires preventing Lost
Update. Accordingly, HATs cannot provide compliant TPC-C execution but
can maintain uniqueness constraints. Second, the New-Order transaction
decrements inventory counts: what if the count becomes negative?
Fortunately, New-Order restocks each item's inventory count
(increments by 91) if it would become negative as the result of
placing an order. This means that, even in the presence of concurrent
New-Orders, an item's stock will never fall below zero. This is TPC-C
compliant, but a HAT system might end up with more stock than in an
unavailable implementation with synchronous coordination.

\vspace{.5em}\noindent\textit{TPC-C Non-monotonicity.} The Delivery
transaction is challenging due to non-monotonicity. Each Delivery
deletes a pending order from the New-Order table and should be
idempotent in order to avoid billing a customer twice; this implies a
need to prevent Lost Update. This issue can be avoided by moving the
non-monotonicity to the real world---the carrier that picks up the
package for an order can ensure that no other carrier will do so---but
cannot provide a correct execution with HATs alone. However, according
to distributed transaction architects~\cite{entitygroup}, these
compensatory actions are relatively common in real-world business
processes.

\vspace{.5em}\noindent\textit{Integrity Constraints.} Throughout execution, TPC-C also requires the maintenance of several
integrity constraints. For example, Consistency Condition 1 (3.3.2.1)
requires that each warehouse's sales count must reflect the sum of its
subordinate sales districts. This integrity constraint spans two
tables but, given the ability to update rows in both tables atomically
via MAV, can be easily maintained. Consistency Conditions 4 through 12
(3.3.2.4-12) can similarly be satisfied by applying updates atomically
across tables. Consistency Conditions 2 and 3 (3.3.2.2-3) concern
order ID assignment and are problematic. Finally, while TPC-C is not
subject to multi-key anomalies, we note that many TPC-E isolation
tests (i.e., simultaneously modifying a product description and its
thumbnail) are also achievable using HATs.

\vspace{.5em}\noindent\textit{Summary.} Many---but not all---TPC-C
transactions are well served by HATs. The two problematic
transactions, New-Order and Payment, rely on non-monotonic state
update. The former can be modified to ensure ID uniqueness but not
sequential ID ordering, while the latter is inherently non-monotonic,
requiring external compensation or stronger consistency
protocols. Based on these experiences and discussions with
practitioners, we believe that HAT guarantees can provide useful
semantics for a large class of application functionality, while a
(possibly small) subset of operations will require stronger,
unavailable properties.

\subsection{Experimental Costs}
\label{sec:prototype}

To demonstrate the performance implications of HAT guarantees in a
real-world environment, we implemented a HAT database prototype. We
verify that, as Section~\ref{sec:latency}'s measurements suggested,
``strongly consistent'' algorithms incur substantial latency penalties
(over WAN, 10 to 100 times higher than their HAT counterparts)
compared to HAT-compliant algorithms, which scale linearly. Our goal
is \textit{not} a complete performance analysis of HAT semantics but
instead a proof-of-concept demonstration of a small subset of the HAT
space on real-world infrastructure.

\vspace{.5em}\noindent\textit{Implementation.} Our prototype database
is a partially replicated (hash-based partitioned) key-value backed by
LevelDB and implemented in Java using Apache Thrift. It currently
supports eventual consistency (hereafter, \texttt{eventual};
last-writer-wins RU with standard all-to-all anti-entropy between
replicas) and the efficient HAT MAV algorithm as sketched in
Section~\ref{sec:ta}. (hereafter, \texttt{MAV}). We support non-HAT
operation whereby all operations for a given key are routed to a
(randomly) designated \texttt{master} replica for each key
(guaranteeing single-key linearizability, as in Gilbert and Lynch's
CAP Theorem proof~\cite{gilbert-cap} and in PNUTS~\cite{pnuts}'s
``read latest'' operation; hereafter, \texttt{master}) as well as
distributed two-phase locking. Servers are durable: they synchronously
write to LevelDB before responding to client requests, while new
writes in MAV are synchronously flushed to a disk-resident write-ahead
log.

\vspace{.5em}\noindent\textit{Configuration.} We deploy the database
in \textit{clusters}---disjoint sets of database servers that each
contain a single, fully replicated copy of the data---typically across
datacenters and stick all clients within a datacenter to their
respective cluster (trivially providing read-your-writes and monotonic
reads guarantees). By default, we deploy 5 Amazon EC2
\texttt{m1.xlarge} instances as servers in each cluster. For our
workload, we link our client library to the YCSB
benchmark~\cite{ycsb}, which is well suited to LevelDB's key-value
schema, grouping every eight YCSB operations from the default workload
(50\% reads, 50\% writes) to form a transaction. We increase the
number of keys in the workload from the default 1,000 to 100,000 with
uniform random key access, keeping the default value size of $1KB$,
and running YCSB for 180 seconds per configuration.

\begin{figure}[t!]
\begin{center}
\hspace{2em}\includegraphics[width=.8\columnwidth]{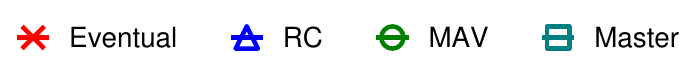}
\end{center}\vspace{-2.5em}
\begin{center}\small\textbf{A.) Within \texttt{us-east} \texttt{(VA)}}\end{center}\vspace{-1.5em}
\includegraphics[width=\figfactor\columnwidth]{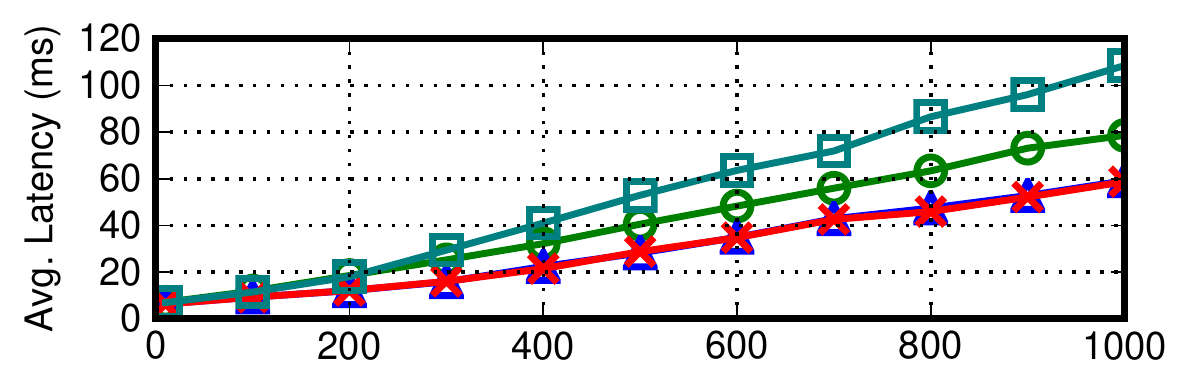}\vspace{-.5em}
\includegraphics[width=\figfactor\columnwidth]{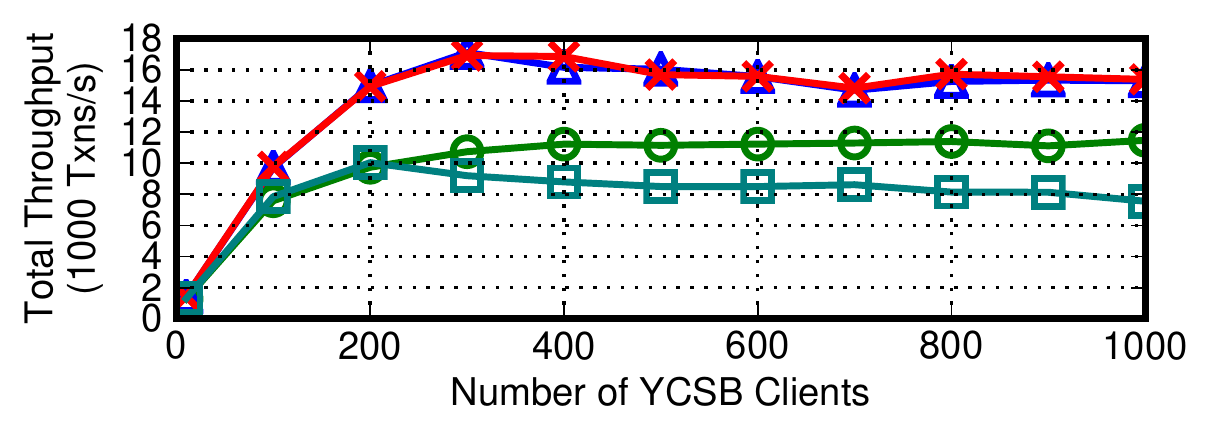}\vspace{-.75em}
\begin{center}\small\textbf{B.) Between \texttt{us-east} \texttt{(CA)} and \texttt{us-west-2} \texttt{(OR)}}\end{center}\vspace{-1.5em}
\includegraphics[width=\figfactor\columnwidth]{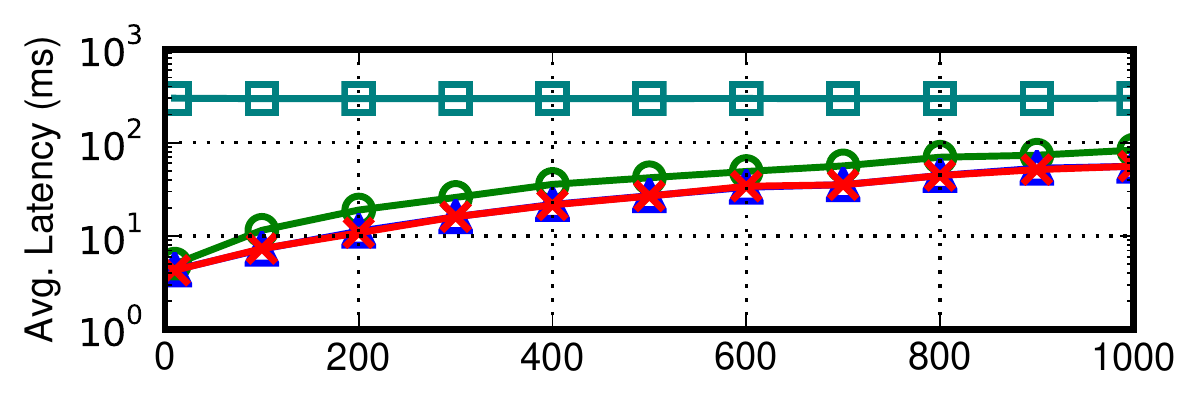}\vspace{-.5em}
\includegraphics[width=\figfactor\columnwidth]{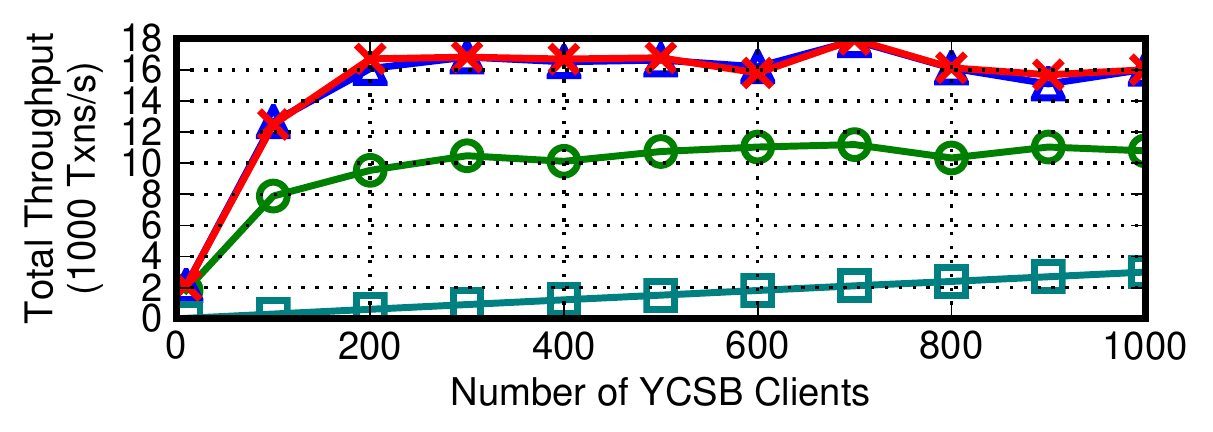}\vspace{-.75em}
\begin{center}\small \textbf{C.) Between}{ \texttt{us-east} \texttt{(VA)}, \texttt{us-west-1} \texttt{(CA)},\\ \texttt{us-west-2} \texttt{(OR)}, \texttt{eu-west} \texttt{(IR)}, \texttt{ap-northeast} \texttt{(SI)}}\end{center}\vspace{-1.5em}
\includegraphics[width=\figfactor\columnwidth]{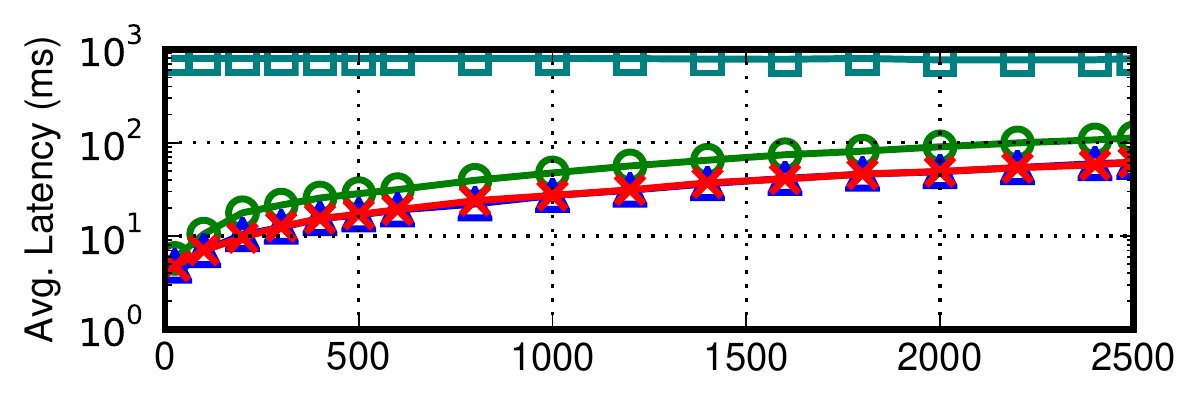}\vspace{--.5em}
\includegraphics[width=\figfactor\columnwidth]{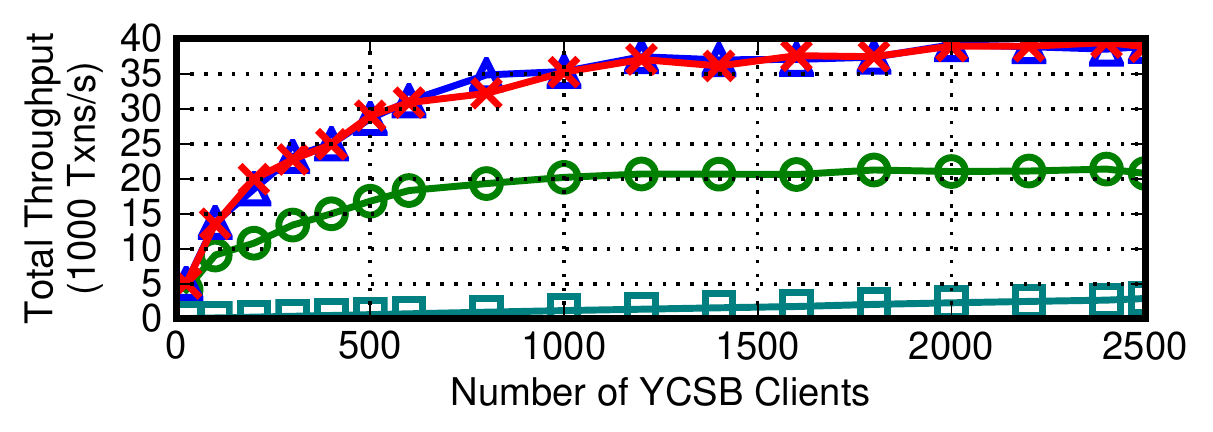}
\caption{YCSB performance for two clusters of five servers each
  deployed within a single datacenter and cross-datacenters.}\vspace{-1.5em}
\label{fig:wan-exp}
\end{figure}

\vspace{.5em}\noindent\textit{Geo-replication.} We first deploy the
database prototype across an increasing number of
datacenters. Figure~\ref{fig:wan-exp}A shows that, when operating two
clusters within a single datacenter, mastering each data item results
in approximately half the throughput and double the latency of
\texttt{eventual}. This is because HAT models are able to utilize
replicas in both clusters instead of always contacting the (single)
master. \texttt{RC}---essentially \texttt{eventual} with
buffering---is almost identical to \texttt{eventual}, while
\texttt{MAV}---which incurs two writes for every client-side write
(i.e., new writes are sent to the WAL then subsequently moved into
LevelDB once stable)---achieves ~75\% of the throughput. Latency
increases linearly with the number of YCSB clients due to contention
within LevelDB.

In contrast, when the two clusters are deployed across the continental
United States (Figure~\ref{fig:wan-exp}B), the average latency of
\texttt{master} increases to $300$ms (a $278$--$4257\%$ latency
increase; average $37$ms latency per operation). For the same number
of YCSB client threads, \texttt{master} has substantially lower
throughput than the HAT configurations. Increasing the number of YCSB
clients \textit{does} increase the throughput of \texttt{master}, but
our Thrift-based server-side connection processing did not gracefully
handle more than several thousand concurrent connections. In contrast,
across two datacenters, the performance of eventual, RC, and MAV are
near identical to a single-datacenter deployment.

When five clusters (as opposed to two, as before) are deployed across
the five EC2 datacenters with lowest communication cost
(Figure~\ref{fig:wan-exp}C), the trend continues: \texttt{master}
latency increases to nearly $800$ms per transaction. As an attempt at
reducing this overhead, we implemented and benchmarked a variant of
quorum-based replication as in Dynamo~\cite{dynamo}, where clients
sent requests to all replicas, which completed as soon as a majority
of servers responded (guaranteeing regular
semantics~\cite{herlihy-art}); this strategy (not pictured) did not
substantially improve performance due to the network topology and
because worst-case server load was unaffected. With five clusters,
\texttt{MAV}'s relative throughput decreased: every YCSB \texttt{put}
operation resulted in four \texttt{put} operations on remote replicas
and, accordingly, the cost of anti-entropy increased (e.g., each
server processed four replicas' anti-entropy as opposed to one before,
reducing the opportunity for batching and decreasing available
resources for incoming client requests). This in turn increased
garbage collection activity and, more importantly, IOPS when compared
to \texttt{eventual} and \texttt{RC}, causing \texttt{MAV} throughput
to peak at around half of \texttt{eventual}. With in-memory
persistence (i.e., no LevelDB or WAL), \texttt{MAV} throughput was
within 20\% of \texttt{eventual}.

We have intentionally omitted performance data for two-phase
locking. \texttt{master} performed \textit{far} better than our
textbook implementation, which, in addition to requiring a WAN
round-trip per operation, also incurred substantial overheads due to
mutual exclusion via locking. We expect that, while techniques like
those recently proposed in Calvin~\cite{calvin} can reduce the
overhead of serializable transactions by avoiding locking, our
mastered implementation and the data from Section~\ref{sec:latency}
are reasonable lower bounds on latency.

\begin{figure}[t!]
\begin{center}
\hspace{2em}\includegraphics[width=.8\columnwidth]{figs/strategylegend.pdf}\vspace{-2em}
\end{center}
\begin{center}
\includegraphics[width=\figfactor\columnwidth]{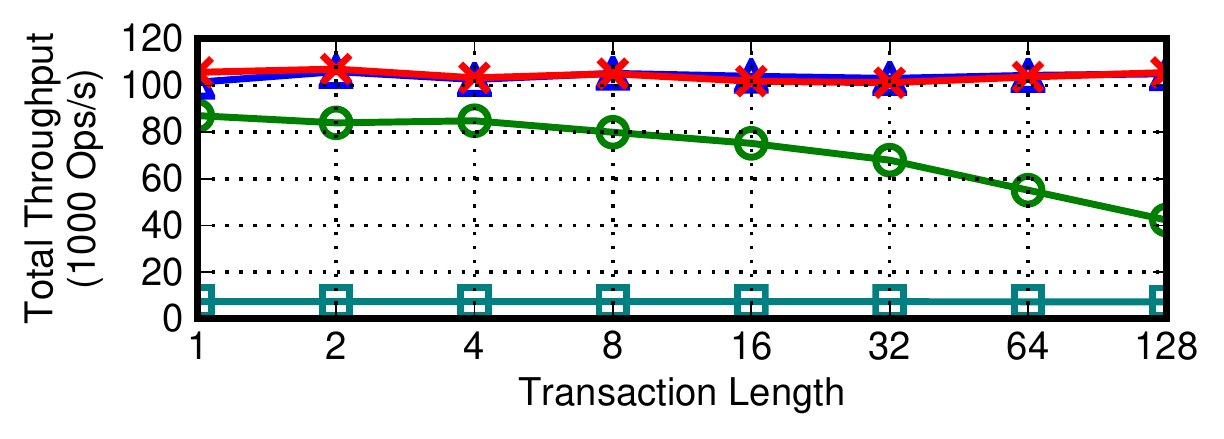}
\end{center}\vspace{-2.25em}
\caption{Transaction length versus throughput.}\vspace{-1em}
\label{fig:txlen}
\begin{center}
\includegraphics[width=\figfactor\columnwidth]{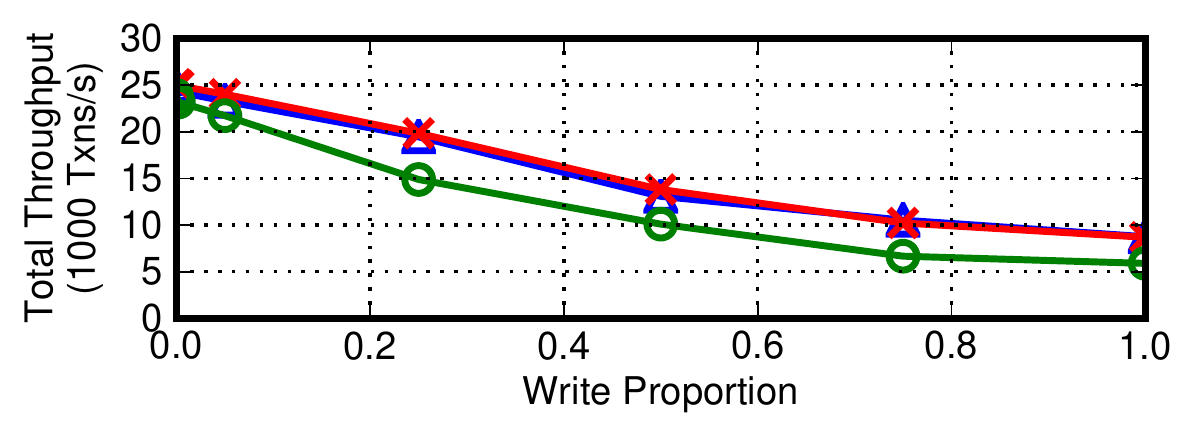}
\end{center}\vspace{-2.25em}
\caption{Proportion of reads and writes versus throughput.}\vspace{-1em}
\label{fig:rprop}
\begin{center}
\includegraphics[width=\figfactor\columnwidth]{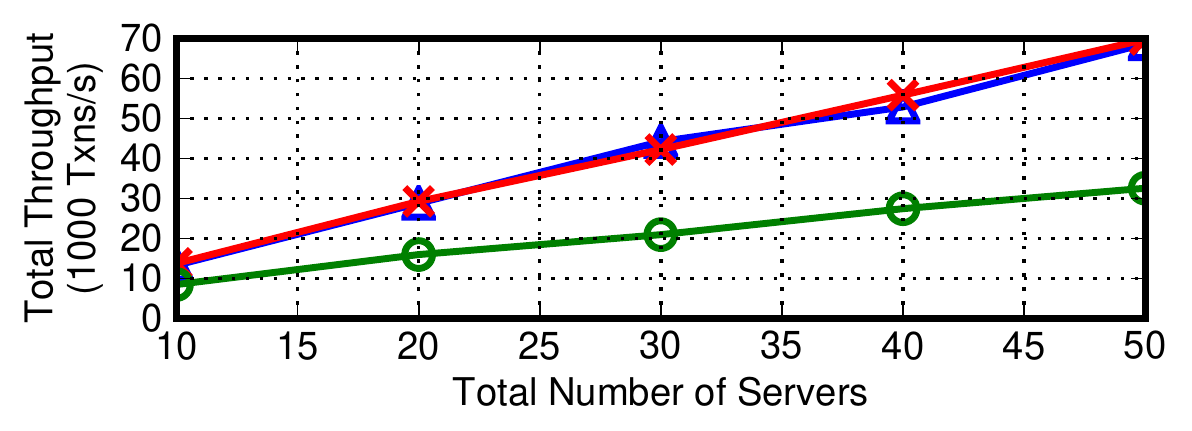}
\end{center}\vspace{-2.25em}
\caption{Scale-out of MAV, Eventual, and RC.}\vspace{-1.5em}
\label{fig:scaleout}
\end{figure}

\vspace{.5em}\noindent\textit{Transaction length.} As shown in
Figure~\ref{fig:txlen} (clusters in Virginia and Oregon), throughput
of \texttt{eventual}, \texttt{RC}, and \texttt{master} operation are
unaffected by transaction length. In contrast, \texttt{MAV} throughput
decreases linearly with increased transaction length: with $1$
operation per transaction, \texttt{MAV} throughput is within 18\% of
\texttt{eventual} ($34$ bytes overhead), and with $128$ operations per
transaction, \texttt{MAV} throughput is within $60\%$ ($1898$ bytes
overhead). This reflects our \texttt{MAV} algorithm's metadata
requirements, which are proportional to transaction length and consume
IOPS and network bandwidth. We are currently investigating alternative
HAT algorithms that do not incur this overhead.

\vspace{.5em}\noindent\textit{Read proportion.} Our default (equal)
proportion of reads and writes is fairly pessimistic: for example,
Facebook reports $99.8\%$ reads for their workload~\cite{eiger}. As
shown in Figure~\ref{fig:rprop} (clusters in Virginia and Oregon),
with all reads, \texttt{MAV} is within $4.8\%$ of \texttt{eventual};
with all writes, \texttt{MAV} is within $33\%$, and the throughput of
\texttt{eventual} decreases by $288.8\%$ compared to all reads. At
$99.8\%$ reads, \texttt{MAV} incurs a $7\%$ overhead ($5.8\%$ for in-memory
storage).

\vspace{.5em}\noindent\textit{Scale-out.} One of the key benefits of
our HAT algorithms is that they are shared-nothing, meaning they
should not compromise scalability. Figure~\ref{fig:scaleout} shows
that varying the number of servers across two clusters in Virginia and
Oregon (with $15$ YCSB clients per server) results in linear scale-out
for \texttt{eventual}, \texttt{RC}, and \texttt{MAV}. \texttt{RC} and
\texttt{eventual} scale linearly: increasing the number of servers per
cluster from $5$ to $25$ yields an approximately $5$x throughput
increase. For the same configurations, \texttt{MAV} scales by $3.8$x,
achieving over $260,000$ operations per second. \texttt{MAV} suffers
from contention in LevelDB---with a memory-backed database,
\texttt{MAV} scales by $4.25$x (not shown)---and MAV-related
performance heterogeneity across servers (Calvin's authors report
similar heterogeneity on EC2~\cite{calvin}). Initial experiments with
a newer prototype including more efficient (non-Thrift) RPC and
connection pooling suggest that this scalability can be substantially
improved.

\vspace{.5em}\noindent\textit{Summary.} Our experimental prototype
confirms our earlier analytical intuitions. HAT systems can provide
useful semantics without substantial performance penalties. In
particular, our MAV algorithm can achieve throughput competitive with
eventual consistency at the expense of increased disk and network
utilization. Perhaps more importantly, all HAT algorithms circumvent
high WAN latencies inevitable with non-HAT implementations. Our results
highlight Deutsch's observation that ignoring factors such as latency
can ``cause big trouble and painful learning
experiences''~\cite{fallacies-deutsch}---in a single-site context,
paying the cost of coordination may be tenable, but, especially as
services are geo-replicated, costs increase.

\section{Related Work}
\label{sec:relatedwork}

We have discussed traditional mechanisms for distributed
coordination and several related systems in
Section~\ref{sec:eval-existing}, but, in this section, we further discuss
related work. In particular, we discuss related work on highly
available semantics, mechanisms for concurrency control, and
techniques for scalable distributed operations.

Weak consistency and high availability have been well
studied. Serializability has long been known to be
unachievable~\cite{davidson-survey} and Brewer's CAP Theorem has
attracted considerable attention~\cite{gilbert-cap}. Recent work on
PACELC expands CAP by considering connections between ``weak
consistency'' and low latency~\cite{abadi-pacelc}, while several
studies examine weak isolation guarantees~\cite{adya,
  ansicritique}. There are a wide range of coordination-avoiding
``optimistic replication'' strategies~\cite{optimistic} and several
recent efforts at further understanding these strategies in light of
current practice and the proliferation of ``eventually consistent''
stores~\cite{bailis-ec, bernstein-survey}. Notably, Bernstein and
Das~\cite{bernstein-survey} specifically highlight the importance of
stickiness~\cite{sessionguarantees, vogels-defs}---which we formalize
in Section~\ref{sec:sticky}.  Aside from our earlier workshop paper
discussing transactional availability, real-world ACID, and HAT RC and
I-CI~\cite{hat-hotos}---which this work expands with additional
semantics, algorithms, and analysis---we believe this paper is the
first to explore the connections between transactional semantics, data
consistency, and (Gilbert and Lynch~\cite{gilbert-cap}) availability.


There has been a recent resurgence of interest in distributed
multi-object semantics, both in academia~\cite{kraska-s3, gstore,
  eiger, walter,calvin, swift} and industry~\cite{orleans,spanner}. As
discussed in Section~\ref{sec:modernacid}, classic ACID databases
provide strong semantics but their lock-based and traditional
multi-versioned implementations are unavailable in the presence of
partitions~\cite{bernstein-book, gray-isolation}. Notably, Google's
Spanner provides strong one-copy serializable transactions. While
Spanner is highly specialized for Google's read-heavy workload, it
relies on two-phase commit and two-phase locking for read/write
transactions~\cite{spanner}. As we have discussed, the penalties
associated with this design are fundamental to serializability. For
users willing to tolerate unavailability and increased latency,
Spanner, or similar ``strongly consistent''
systems~\cite{kemme-classification}---including Calvin~\cite{calvin},
G-Store~\cite{gstore}, HBase, HStore~\cite{hstore},
Orleans~\cite{orleans}, Postgres-R~\cite{kemme-thesis},
Walter~\cite{walter}, and a range of snapshot isolation
techniques~\cite{daudjee-session}---reasonable
choices.

With HATs, we seek an alternative set of transactional semantics that
are still useful but do not violate requirements for high availability
or low latency. Recent systems proposals such as Swift~\cite{swift},
Eiger~\cite{eiger}, and Bolt-on Causal Consistency~\cite{bolton}
provide transactional causal consistency guarantees with varying
availability and represent a new class of sticky HAT systems. There
are infinitely many HAT models (i.e., always reading value 1 is
incomparable with always returning value 2), but a recent report from
UT Austin shows that no model stronger than causal consistency is
achievable in a sticky highly available, \textit{one-way convergent}
system~\cite{cac}. This result is promising and complementary to our
results for general-purpose convergent data stores. Finally, Burkhardt
et al. have concurrently developed an axiomatic specification for
eventual consistency; their work-in-progress report contains alternate
formalism for several HAT guarantees~\cite{burkhardt-txns}.




\section{Conclusions and Future Work}
\label{sec:conclusion}

The current state of database software offers uncomfortable and
unnecessary choices between availability and transactional semantics.
Through our analysis and experiments, we have demonstrated how goals
of high availability will remain a critical aspect of many future data
storage systems. We expose a broad design space of Highly Available
Transactions (HATs), which can offer the key benefits of highly
available distributed systems---``always on'' operation during
partitions and low-latency operations (often orders of magnitude lower
than non-compliant)---while also providing a family of transactional
isolation levels and replica semantics that have been adopted in
practice.  We also identify many semantic guarantees that are
unachievable with high availability, including Lost Update and Write
Skew anomaly prevention, concurrent update prevention, and bounds on
data recency. Despite these limitations, and somewhat surprisingly,
many of the default (and sometimes strongest) semantics provided by
today's traditional database systems are achievable as HATs, hinting
that distributed databases need not compromise availability, low
latency, or scalability in order to serve many existing applications.

In this paper, we have largely focused on previously defined isolation
and data consistency models from the consistency from the database and
distributed systems communities. Their previous definitions and, in
many cases, widespread adoption hints at their utility to
end-users. However, we believe there is considerable work to be done
to improve the programmability of highly available systems. Isolation
and data consistency are means by which \textit{application-level}
consistency is achieved but are typically not end goals for end-user
applications. Our results hint that a mix of HAT and non-HAT semantics
(the latter used sparingly) is required for practical applications,
but the decision to employ each and the system architectures for a
hybrid approach remain open problems. While we have studied the
analytical and experiment behavior of several HAT models, there is
substantial work in further understanding the performance and design
of systems within the large set of HAT models. Weakened failure
assumptions as in escrow or in the form of bounded network asynchrony
could enable richer HAT semantics at the cost of general-purpose
availability. Alternatively, there is a range of possible solutions
providing strong semantics during partition-free periods and weakened
semantics during partitions. Based on our understanding of what is
desired in the field and newfound knowledge of what is possible to
achieve, we believe HATs represent a large and useful design space for
exploration.

\vspace{.5em}\noindent\textbf{Acknowledgments} We would like to thank
Peter Alvaro, Neil Conway, Evan Jones, Adam Oliner, Aurojit Panda,
Shivaram Venkataraman, and the HotOS and VLDB reviewers for their
helpful feedback on this work. This research was supported in part by
the Air Force Office of Scientiﬁc Research (grant FA95500810352),
DARPA XData Award FA8750-12-2-0331, the National Science Foundation
(grants CNS-0722077, IIS-0713661, and IIS-0803690), NSF CISE
Expeditions award CCF-1139158, the National Science Foundation
Graduate Research Fellowship (grant DGE-1106400), and by gifts from
Amazon Web Services, Google, SAP, Cisco, Clearstory Data, Cloudera,
Ericsson, Facebook, FitWave, General Electric, Hortonworks, Huawei,
Intel, Microsoft, NetApp, Oracle, Samsung, Splunk, VMware, WANdisco,
and Yahoo!.\vspace{.5em}

\bibliographystyle{abbrv}
\scriptsize

\bibliography{hat-vldb}

\begin{thebibliography}{10}

\bibitem{abadi-pacelc}
D.~J. Abadi.
\newblock Consistency tradeoffs in modern distributed database system design:
  {CAP} is only part of the story.
\newblock IEEE Computer, 45(2), 2012.

\bibitem{adya}
A.~Adya.
\newblock {\em Weak consistency: a generalized theory and optimistic
  implementations for distributed transactions}.
\newblock PhD thesis, MIT, 1999.

\bibitem{causalmemory}
M.~Ahamad, G.~Neiger, J.~E. Burns, P.~Kohli, and P.~Hutto.
\newblock Causal memory: Definitions, implementation and programming.
\newblock {\em Dist. Comp.}, 9(1), 1995.

\bibitem{calm}
P.~Alvaro, N.~Conway, J.~M. Hellerstein, and W.~R. Marczak.
\newblock Consistency analysis in {Bloom}: a {CALM} and collected approach.
\newblock In {\em CIDR 2011}.

\bibitem{ansi-sql}
{ISO/IEC 9075-2:2011} \textit{Information technology -- Database languages --
  SQL -- Part 2: Foundation (SQL/Foundation)}.

\bibitem{1sr}
R.~Attar, P.~A. Bernstein, and N.~Goodman.
\newblock Site initialization, recovery, and backup in a distributed database
  system.
\newblock {\em IEEE Trans. Softw. Eng.}, 10(6):645--650, Nov. 1984.

\bibitem{amazon-netpartition}
{AWS}.
\newblock {Summary of the Amazon EC2 and Amazon RDS Service Disruption in the
  US East Region}.
\newblock \url{http://tinyurl.com/6ab6el6}, April 2011.

\bibitem{hat-hotos}
P.~Bailis, A.~Fekete, A.~Ghodsi, J.~M. Hellerstein, and I.~Stoica.
\newblock {HAT, not CAP}: Introducing {Highly Available Transactions}.
\newblock In {\em HotOS 2013}.

\bibitem{bailis-ec}
P.~Bailis and A.~Ghodsi.
\newblock {Eventual Consistency} today: Limitations, extensions, and beyond.
\newblock {\em ACM Queue}, 11(3), March 2013.

\bibitem{bolton}
P.~Bailis, A.~Ghodsi, J.~M. Hellerstein, and I.~Stoica.
\newblock Bolt-on causal consistency.
\newblock In {\em SIGMOD 2013}.

\bibitem{ansicritique}
H.~Berenson, P.~Bernstein, J.~Gray, J.~Melton, E.~O'Neil, and P.~O'Neil.
\newblock A critique of {ANSI SQL} isolation levels.
\newblock In {\em SIGMOD 1995}.

\bibitem{bernstein-survey}
P.~Bernstein and S.~Das.
\newblock Rethinking eventual consistency.
\newblock In {\em SIGMOD}, 2013.

\bibitem{bernstein-book}
P.~Bernstein, V.~Hadzilacos, and N.~Goodman.
\newblock {\em Concurrency control and recovery in database systems}, volume
  370.
\newblock Addison-wesley New York, 1987.

\bibitem{kraska-s3}
M.~Brantner, D.~Florescu, D.~Graf, D.~Kossmann, and T.~Kraska.
\newblock Building a database on {S3}.
\newblock In {\em SIGMOD 2008}.

\bibitem{brewer-slides}
E.~Brewer.
\newblock Towards robust distributed systems.
\newblock {Keynote at PODC 2000}.

\bibitem{sessiontocausal}
J.~Brzezinski, C.~Sobaniec, and D.~Wawrzyniak.
\newblock From session causality to causal consistency.
\newblock In {\em PDP 2004}.

\bibitem{burkhardt-txns}
S.~Burckhardt, A.~Gotsman, and H.~Yang.
\newblock Understanding eventual consistency.
\newblock Technical Report MSR-TR-2013-39.
\newblock \url{http://tinyurl.com/bqty9yz}.

\bibitem{orleans}
S.~Bykov, A.~Geller, G.~Kliot, J.~R. Larus, R.~Pandya, and J.~Thelin.
\newblock Orleans: cloud computing for everyone.
\newblock In {\em SOCC 2011}.

\bibitem{readonly}
A.~Chan and R.~Gray.
\newblock Implementing distributed read-only transactions.
\newblock {\em IEEE Transactions on Software Engineering}, (2):205--212, 1985.

\bibitem{bigtable}
F.~Chang, J.~Dean, S.~Ghemawat, et~al.
\newblock Bigtable: A distributed storage system for structured data.
\newblock In {\em OSDI 2006}.

\bibitem{foundation-article}
G.~Clarke.
\newblock {The Register: NoSQL's CAP theorem busters: We don't drop ACID}.
\newblock \url{http://tinyurl.com/bpsug4b}, November 2012.

\bibitem{pnuts}
B.~Cooper et~al.
\newblock {PNUTS}: {Yahoo!'s} hosted data serving platform.
\newblock In {\em VLDB 2008}.

\bibitem{ycsb}
B.~F. Cooper, A.~Silberstein, E.~Tam, R.~Ramakrishnan, and R.~Sears.
\newblock Benchmarking cloud serving systems with {YCSB}.
\newblock In {\em ACM SOCC 2010}.

\bibitem{spanner}
J.~C. Corbett, J.~Dean, M.~Epstein, A.~Fikes, C.~Frost, J.~J. Furman, et~al.
\newblock Spanner: Google's globally-distributed database.
\newblock In {\em OSDI 2012}.

\bibitem{gstore}
S.~Das, D.~Agrawal, and A.~El~Abbadi.
\newblock G-store: a scalable data store for transactional multi key access in
  the cloud.
\newblock In {\em SOCC 2010}, pages 163--174.

\bibitem{daudjee-session}
K.~Daudjee and K.~Salem.
\newblock Lazy database replication with ordering guarantees.
\newblock In {\em ICDE 2004}, pages 424--435.

\bibitem{davidson-survey}
S.~Davidson, H.~Garcia-Molina, and D.~Skeen.
\newblock Consistency in partitioned networks.
\newblock {\em ACM CSUR}, 17(3):341--370, 1985.

\bibitem{dean-keynote}
J.~Dean.
\newblock Designs, lessons and advice from building large distributed systems.
\newblock Keynote at LADIS 2009.

\bibitem{dynamo}
G.~DeCandia, D.~Hastorun, M.~Jampani, G.~Kakulapati, A.~Lakshman, et~al.
\newblock Dynamo: {Amazon's} highly available key-value store.
\newblock In {\em SOSP 2007}.

\bibitem{antientropy}
A.~Demers et~al.
\newblock Epidemic algorithms for replicated database maintenance.
\newblock In {\em PODC}, 1987.

\bibitem{fallacies-deutsch}
P.~Deutsch.
\newblock The eight fallacies of distributed computing.
\newblock \url{http://tinyurl.com/c6vvtzg}, 1994.

\bibitem{ec2-downsites}
R.~Dillet.
\newblock Update: {Amazon Web Services} down in {North Virginia} — {Reddit,
  Pinterest, Airbnb, Foursquare, Minecraft} and others affected.
\newblock TechCrunch \url{http://tinyurl.com/9r43dwt}, October 2012.

\bibitem{snapshot-serializable}
A.~Fekete, D.~Liarokapis, E.~O'Neil, P.~O'Neil, and D.~Shasha.
\newblock Making snapshot isolation serializable.
\newblock {\em ACM TODS}, 30(2):492--528, June 2005.

\bibitem{gilbert-cap}
S.~Gilbert and N.~Lynch.
\newblock Brewer's conjecture and the feasibility of consistent, available,
  partition-tolerant web services.
\newblock {\em SIGACT News}, 33(2):51--59, 2002.

\bibitem{sigcomm-dc}
P.~Gill, N.~Jain, and N.~Nagappan.
\newblock Understanding network failures in data centers: measurement,
  analysis, and implications.
\newblock In {\em SIGCOMM 2011}.

\bibitem{gray-virtues}
J.~Gray.
\newblock The transaction concept: Virtues and limitations.
\newblock In {\em VLDB 1981}.

\bibitem{gray-isolation}
J.~Gray, R.~Lorie, G.~Putzolu, and I.~Traiger.
\newblock Granularity of locks and degrees of consistency in a shared data
  base.
\newblock Technical report, IBM, 1976.

\bibitem{hamilton-partitions}
J.~Hamilton.
\newblock {Stonebraker on CAP Theorem and Databases}.
\newblock \url{http://tinyurl.com/d3gtfq9}, April 2010.

\bibitem{entitygroup}
P.~Helland.
\newblock Life beyond distributed transactions: an apostate's opinion.
\newblock In {\em CIDR 2007}.

\bibitem{herlihy-art}
M.~Herlihy and N.~Shavit.
\newblock {\em The art of multiprocessor programming}.
\newblock 2008.

\bibitem{hstore}
R.~Kallman et~al.
\newblock H-store: a high-performance, distributed main memory transaction
  processing system.
\newblock In {\em VLDB 2008}.

\bibitem{kemme-thesis}
B.~Kemme.
\newblock {\em Database replication for clusters of workstations}.
\newblock PhD thesis, EPFL, 2000.

\bibitem{aphyr-post}
K.~Kingsbury and P.~Bailis.
\newblock The network is reliable.
\newblock June 2013.
\newblock \url{http://aphyr.com/posts/288-the-network-is-reliable}.

\bibitem{labovitz-failures}
C.~Labovitz, A.~Ahuja, and F.~Jahanian.
\newblock Experimental study of internet stability and backbone failures.
\newblock In {\em FTCS 1999}.

\bibitem{lamportclocks}
L.~Lamport.
\newblock Time, clocks, and the ordering of events in a distributed system.
\newblock {\em Commun. ACM}, 21(7):558--565, July 1978.

\bibitem{uw-failure-networks}
V.~Liu, D.~Halperin, A.~Krishnamurthy, and T.~Anderson.
\newblock F10: Fault tolerant engineered networks.
\newblock In {\em NSDI 2013}.

\bibitem{eiger}
W.~Lloyd, M.~J. Freedman, M.~Kaminsky, and D.~G. Andersen.
\newblock Stronger semantics for low-latency geo-replicated storage.
\newblock In {\em NSDI 2013}.

\bibitem{cac}
P.~Mahajan, L.~Alvisi, and M.~Dahlin.
\newblock Consistency, availability, convergence.
\newblock Technical Report TR-11-22, CS Department, UT Austin, May 2011.

\bibitem{ip-backbone-failures}
A.~Markopoulou et~al.
\newblock Characterization of failures in an operational {IP} backbone network.
\newblock {\em IEEE/ACM TON}, 16(4).

\bibitem{pakistan-youtube}
D.~McCullagh.
\newblock How {Pakistan} knocked {YouTube} offline (and how to make sure it
  never happens again).
\newblock CNET, \url{http://tinyurl.com/c4pffd}, February 2008.

\bibitem{research-experiment-partition}
R.~McMillan.
\newblock Research experiment disrupts internet, for some.
\newblock {Computerworld}, \url{http://tinyurl.com/23sqpek}, August 2010.

\bibitem{pnuts-update}
P.~P.~S. Narayan.
\newblock Sherpa update.
\newblock YDN Blog, \url{http://tinyurl.com/c3ljuce}, June 2010.

\bibitem{transaction-liveness}
F.~Pedone and R.~Guerraoui.
\newblock On transaction liveness in replicated databases.
\newblock In {\em Pacific Rim International Symposium on Fault-Tolerant
  Systems}, 1997.

\bibitem{optimistic}
Y.~Saito and M.~Shapiro.
\newblock Optimistic replication.
\newblock {\em ACM Comput. Surv.}, 37(1), Mar. 2005.

\bibitem{deterministic-scheduling}
A.~Schiper and M.~Raynal.
\newblock From group communication to transactions in distributed systems.
\newblock {\em CACM}, 39(4), 1996.

\bibitem{juniper-partition}
L.~Segall.
\newblock Internet routing glitch kicks millions offline.
\newblock CNNMoney, \url{http://tinyurl.com/cmqqac3}, November 2011.

\bibitem{crdt}
M.~Shapiro, N.~Pregui{\c{c}}a, C.~Baquero, and M.~Zawirski.
\newblock A comprehensive study of convergent and commutative replicated data
  types.
\newblock {INRIA TR 7506}, 2011.

\bibitem{walter}
Y.~Sovran, R.~Power, M.~K. Aguilera, and J.~Li.
\newblock Transactional storage for geo-replicated systems.
\newblock In {\em SOSP}, pages 385--400, 2011.

\bibitem{sessionguarantees}
D.~B. Terry, A.~J. Demers, K.~Petersen, M.~J. Spreitzer, M.~M. Theimer, et~al.
\newblock Session guarantees for weakly consistent replicated data.
\newblock In {\em PDIS 1994}.

\bibitem{calvin}
A.~Thomson, T.~Diamond, S.~Weng, K.~Ren, P.~Shao, and D.~Abadi.
\newblock Calvin: Fast distributed transactions for partitioned database
  systems.
\newblock In {\em SIGMOD 2012}.

\bibitem{turner2012failure}
D.~Turner, K.~Levchenko, J.~C. Mogul, S.~Savage, and A.~C. Snoeren.
\newblock On failure in managed enterprise networks.
\newblock HP Labs HPL-2012-101, 2012.

\bibitem{sigcomm-wan}
D.~Turner, K.~Levchenko, A.~C. Snoeren, and S.~Savage.
\newblock California fault lines: understanding the causes and impact of
  network failures.
\newblock {\em SIGCOMM 2011}.

\bibitem{vogels-defs}
W.~Vogels.
\newblock Eventually consistent.
\newblock {\em CACM}, 52(1):40--44, Jan. 2009.

\bibitem{kemme-classification}
M.~Wiesmann, F.~Pedone, A.~Schiper, B.~Kemme, and G.~Alonso.
\newblock Database replication techniques: A three parameter classification.
\newblock In {\em SRDS 2000}.

\bibitem{bobtail}
Y.~Xu, Z.~Musgrave, B.~Noble, and M.~Bailey.
\newblock Bobtail: avoiding long tails in the cloud.
\newblock In {\em NSDI}, 2013.

\bibitem{swift}
M.~Zawirski, A.~Bieniusa, V.~Balegas, N.~Preguica, S.~Duarte, M.~Shapiro, and
  C.~Baquero.
\newblock Geo-replication all the way to the edge.
\newblock Personal communication and draft under submission.
  \url{http://tinyurl.com/cp68svy}.

\end{thebibliography}



\iftechreport
\pagebreak
\begin{appendix}

\normalsize

\section{Formal Definitions}

In this section, we formally define HAT transactional semantics. Our
formalism is based off of that of Adya~\cite{adya}. For the reader
familiar with his formalism, this is a mostly-straightforward exercise
combining transactional models with distributed systems
semantics. While the novel aspects of this work largely pertain
to \textit{using} these definitions (e.g., in Section~\ref{sec:hats}),
we believe it is instructive to accompany them by appropriate
definitions to both eliminate any ambiguity in prose and for the
enjoyment of more theoretically-inclined readers.

\subsection{Model}

Here, we briefly describe our database model. It is, with the
exception of sessions, identical to that of Adya~\cite{adya}. We omit
a full duplication of his formalism here but highlight several salient
criteria. We refer the interested reader to Adya's Ph.D. thesis,
Section 3.1 (pp. 33--43).

Users submit transactions to a database system that contains multiple
versions of sets of objects. Each transaction is composed of writes,
which create new versions of an object, and reads, which return a
written version or the initial version of the object. A transaction's
last operation is either \texttt{commit} or \texttt{abort}, and there
is exactly one invocation of either these two operations per
transaction. Transactions can either read individual items or read
based on predicates, or logical ranges of data items.

\begin{definition}[Version set of a predicate-based operation]
When a transaction executes a read or write based on a predicate P,
the system selects a version for each tuple in P’s relations. The set
of selected versions is called the \textit{Version set} of this predicate-based
operation and is denoted by Vset(P).
\end{definition}

A history over transactions has two parts: a partial order of events
reflects the ordering of operations with respect to each transaction
and a (total) version order ($\ll$) on the committed versions of each
object.

As a departure from Adya's formalism, to capture the use of session
guarantees, we allow transactions to be grouped into sessions. We
represent sessions as a partial ordering on committed transactions
such that each transaction in the history appears in at most one
session.

\subsection{Conflict and Serialization Graphs}

To reason about isolation anomalies, we use Adya's concept of a
conflict graph, which is composed of dependencies between
transaction. The definitions in this section are directly from Adya,
with two differences. First, we expand Adya's formalism to deal
with \textit{per-item} dependencies. Second, we define session
dependencies (Definition~\ref{def:sd1}).

\begin{definition}[Change the Matches of a Predicate-Based Read]
A transaction $T_i$ changes the matches
of a predicate-based read $r_j$(P:Vset(P)) if $T_i$ installs $x_i$, $x_h$
immediately precedes $x_i$ in the version order, and $x_h$ matches
P whereas $x_i$ does not or vice-versa. In this case, we also
say that $x_i$ changes the matches of the predicate-based read.
The above deﬁnition identiﬁes $T_i$ to be a transaction where
a change occurs for the matched set of $r_j$ (P: Vset(P)).
\end{definition}

\begin{definition}[Directly Read-Depends{~\cite[Definition 2]{adya}}]
$T_j$ directly read-depends on transaction $T_i$ if it directly
item-read-depends or directly predicate-read-depends on $T_i$.
\end{definition}

\begin{definition}[Directly item-read-depends by $x$]
$T_j$ directly item-read-depends on transaction $T_i$ if $T_i$ installs some
  object version $x_i$ and $T_j$ reads $x_i$.
\end{definition}

\begin{definition}[Directly item-read-depends]
$T_j$ directly item-read-depends on transaction $T_i$ if $T_j$
  directly item-read-depends by $x$ on $T_i$ for some data item $x$.
\end{definition}

\begin{definition}[Directly predicate-read-depends by $P$]
Transaction $T_j$ directly predicate-read-depends by $P$ on
transaction $T_i$ if $T_j$ performs an operation $r_j$(P: Vset(P)),
$x_k$ $\in$ Vset(P), $i = k$ or $x_i \ll x_k$ , and $x_i$ changes the
matches of $r_j$ (P: Vset(P)).
\end{definition}

\begin{definition}[Directly predicate-read-depends]
$T_j$ directly predicate-read-depends on $T_i$ if $T_j$ directly
  predicate-read-depends by $P$ on $T_i$ for some predicate $P$.
\end{definition}

\begin{definition}[Directly Anti-Depends{~\cite[Definition 4]{adya}}]
Transaction $T_j$ directly anti-depends on transaction $T_i$ if it
directly item-anti-depends or directly predicate-anti-depends on
$T_i$.
\end{definition}

\begin{definition}[Directly item-anti-depends by $x$]
$T_j$ directly item-anti-depends by $x$ on transaction $T_i$ if $T_i$
  reads some object version $x_k$ and $T_j$ installs $x$'s next
  version (after $x_k$) in the version order. Note that the
  transaction that wrote the later version directly item-anti-depends
  on the transaction that read the earlier version.
\end{definition}

\begin{definition}[Directly item-anti-depends]
$T_j$ directly item-anti-depends on transaction $T_i$ if $T_j$
  directly item-anti-depends on transaction $T_i$.
\end{definition}

\begin{definition}[Directly predicate-anti-depends by $P$]
$T_j$ directly predicate-anti-depends by $P$ on transaction $T_i$ if
  $T_j$ overwrites an operation $r_i(P:$ Vset(P)). That is, if $T_j$
  installs a later version of some object that changes the matches of
  a predicate based read performed by $T_i$.
\end{definition}

\begin{definition}[Directly predicate-anti-depends by $P$]
$T_j$ directly predicate-anti-depends on transaction $T_i$ if $T_j$
  directly predicate anti-depends by $P$ on $T_i$ for some predicate
  $P$.
\end{definition}

\begin{definition}[Directly Write-Depends by $x$]
A transaction $T_j$ directly write-depends by $x$ on transaction $T_i$
if $T_i$ installs a version $x_i$ and $T_j$ installs $x$'s next
version (after $x_i$) in the version order.
\end{definition}

\begin{definition}[Directly Write-Depends{~\cite[Definition 5]{adya}}]
A transaction $T_j$ directly write-depends on transaction $T_i$ if
$T_i$ directly write-depends by $x$ on $T_j$ for some item $x$.
\end{definition}

\begin{definition}[Session-Depends]
\label{def:sd1}
A transaction $T_j$ session-depends on transaction $T_i$ if $T_i$ and
$T_j$ occur in the same session and $T_i$ precedes $T_j$ in the
session commit order.
\end{definition}

The dependencies for a history $H$ form a graph called its Directed
Serialization Graph ($DSG(H)$). If $T_j$ directly write-depends on
$T_i$ by $x$, we draw $T_i\overset{ww_x}\longrightarrow T_j$. If $T_j$
read-depends on $T_i$ by $x$, we draw
$T_i \overset{rw_x}\longrightarrow T_j$. If $T_j$ directly
anti-depends on transaction $T_j$ by $x$, we draw
$T_i \overset{wr_x}\DashedArrow T_j$. If $T_j$ session-depends on
$T_i$ in session $S$, we draw $T_i \overset{S}\longrightarrow
T_j$~\cite[Definition 8]{adya}.

We also consider the Unfolded Serialization Graph ($USG(H)$) that is a
variation of the $DSG$. The USG is speciﬁed for the transaction of
interest, $T_i$, and a history, $H$, and is denoted by $USG(H,
T_i)$. For the USG, we retain all nodes and edges of the $DSG$ except
for $T_i$ and the edges incident on it. Instead, we split the node for
$T_i$ into multiple nodes---one node for every read/write event in
$T_i$. The edges are now incident on the relevant event of $T_i$.

$USG(H, T_i)$ is obtained by transforming $DSG(H)$ as follows:.  For
each node $p$ ($p \neq T_i$) in $DSG(H)$, we add a node to
$USG(H,T_i)$. For each edge from node $p$ to node $q$ in $DSG(H)$,
where p and q are different from $T_i$, we draw a corresponding edge
in $USG(H,T_i)$. Now we add a node corresponding to every read and
write performed by $T_i$. Any edge that was incident on $T_i$ in the
$DSG$ is now incident on the relevant event of $T_i$ in the
$USG$. Finally, consecutive events in $T_i$ are connected
by \textit{order edges}, e.g., if an action (e.g., SQL statement)
reads object $y_j$ and immediately follows a write on object $x$ in
transaction $T_i$, we add an order-edge from $w_i(x_i)$ to
$r_i(y_j)$~\cite[Section 4.2.1]{adya}.

\subsection{Transactional Anomalies and Isolation Levels}

Following Adya, we define isolation levels according to
possible \textit{anomalies}--typically represented by cycles in the
serialization graphs. Definitions~\ref{def:n-ici}--\ref{def:lostupdate}
are not found in Adya but are found (albeit not in this formalism) in
Berenson et al.~\cite{ansicritique} and the literature on session
guarantees~\cite{sessionguarantees, vogels-defs}.

\begin{definition}[Write Cycles (G0)]
A history $H$ exhibits phenomenon G0 if $DSG(H)$ contains a directed
cycle consisting entirely of write-dependency edges.
\end{definition}

\begin{definition}[Read Uncommitted]
A system that provides Read Uncommitted isolation prohibits phenomenon G0.
\end{definition}

\begin{definition}[Aborted Reads (G1a)]
A history $H$ exhibits phenomenon G1a if it contains an aborted
transaction $T_1$ and a committed transaction $T_2$ such that $T_2$ has read
some object (maybe via a predicate) modiﬁed by $T_1$.
\end{definition}

\begin{definition}[Intermediate Reads (G1b)]
A history $H$ exhibits phenomenon G1b if it contains a committed
transaction $T_2$ that has read a version of object $x$ (maybe via a
predicate) written by transaction $T_1$ that was not $T_1$’s ﬁnal
modiﬁcation of $x$.
\end{definition}

\begin{definition}[Circular Information Flow (G1c)]
A history $H$ exhibits phenomenon G1c if $DSG(H)$ contains a directed
cycle consisting entirely of dependency edges.
\end{definition}

\begin{definition}[Read Committed]
A system that provides Read Committed isolation prohibits phenomenon G0, G1a, G1b, and G1c.
\end{definition}

\begin{definition}[Item-Many-Preceders (IMP)]
\label{def:imp}
A history $H$ exhibits phenomenon IMP if $DSG(H)$ contains a
transaction $T_i$ such that $T_i$ directly item-read-depends by $x$ on more
than one other transaction.
\end{definition}

\begin{figure}[H]
\begin{align*}
\small
T_1 &: w_x(1)\\
T_2 &: w_x(2)\\
T_3 &: r_x(1)~r_x(2)
\end{align*}
\caption{Example of \textit{IMP} anomaly.}
\label{fig:nici-history}
\end{figure}

\begin{figure}[H]
\centering
\begin{tikzpicture}[node distance=3cm]
\tikzstyle{tx}=[draw=none, fill=none]
\node[tx] (T1) at (0,0) {$T_1$};
\node[tx] (T2) at (0,-1) {$T_2$};
\node[tx] (T3) at (2,-.5) {$T_3$};

\draw[->] (T1) edge node[above]{$wr_x$} (T3);
\draw[->] (T2) edge node[below]{$wr_x$} (T3);
\end{tikzpicture}
\caption{DSG for Figure~\ref{fig:nici-history}.}
\label{fig:nici-dsg}
\end{figure}
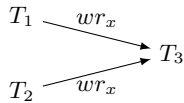

\begin{definition}[Item Cut Isolation (I-CI)]
A system that provides Item Cut Isolation prohibits phenomenon IMP.
\end{definition}

\begin{definition}[Predicate-Many-Preceders (PMP)]
A history $H$ exhibits phenomenon PMP if, for all predicate-based
reads $r_i(P_i:Vset(P_i))$ and $r_j(P_j:Vset(P_j)$ in $T_k$ such that
the logical ranges of $P_i$ and $P_j$ overlap (call it $P_o$), the set
of transactions that change the matches of $P_o$ for $r_i$ and $r_j$
differ.
\end{definition}

\begin{definition}[Predicate Cut Isolation (P-CI)]
A system that provides Predicate Cut Isolation prohibits phenomenon PMP.
\end{definition}

\begin{definition}[Observed Transaction Vanishes (OTV)]
A history $H$ exhibits phenomenon OTV if $USG(H)$ contains a directed
cycle consisting of exactly one read-dependency edge by $x$ from $T_j$
to $T_i$ and a set of edges by $y$ containing at least one
anti-dependency edge from $T_i$ to $T_j$ and $T_j$'s read from $y$
precedes its read from $x$.
\end{definition}

\begin{figure}[H]
\begin{align*}
\small
T_1 &: w_x(1)~w_y(1)\\
T_2 &: w_x(2)~w_y(2)\\
T_3 &: r_x(2)~r_y(1)
\end{align*}
\caption{Example of \textit{OTV} anomaly.}
\label{fig:nta-history}
\end{figure}

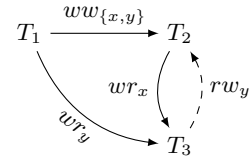
\begin{figure}[H]
\centering
\begin{tikzpicture}[node distance=3cm]
\tikzstyle{tx}=[draw=none, fill=none]
\node[tx] (T1) at (0,0) {$T_1$};
\node[tx] (T2) at (2,0) {$T_2$};
\node[tx] (T3) at (2,-1.5) {$T_3$};

\draw[->] (T1) edge node[sloped, above]{$ww_{\{x, y\}}$} (T2);
\draw[->] (T1) edge [bend right] node[sloped, below]{$wr_y$} (T3);
\draw[->] (T2) edge [bend right] node[left]{$wr_x$} (T3);
\draw[dashed, ->] (T3) edge [bend right] node[right]{$rw_y$} (T2);
\end{tikzpicture}
\caption{DSG for Figure~\ref{fig:nta-history}.}
\label{fig:nta-dsg}
\end{figure}

\begin{definition}[Monotonic Atomic View (MAV)]
A system that provides Monotonic Atomic View isolation prohibits
phenomenon OTV in addition to providing Read Committed isolation.
\end{definition}

The following session guarantees are directly adapted from Terry et
al.'s original definitions~\cite{sessionguarantees}:

\begin{definition}[Non-monotonic Reads (N-MR)]
A history $H$ exhibits phenomenon N-MR if $DSG(H)$ contains a directed cycle
consisting of a transitive session-dependency between transactions
$T_j$ and $T_i$ with an anti-dependency edge by $i$ from $T_j$ and a
read-dependency edge by $i$ into $T_i$.
\end{definition}

\begin{definition}[Monotonic Reads (MR)]
A system provides Monotonic Reads if it prohibits phenomenon N-MR.
\end{definition}

\begin{figure}[H]
\begin{align*}
\small
T_1 &: w_x(1)\\
T_2 &: w_x(2)\\
T_3 &: r_x(2)\\
T_4 &: r_x(1)
\end{align*}
\caption{Example of \textit{N-MR} violation when $w_x(1) \ll w_x(2)$ and $T_4$ directly session-depends on $T_3$.}
\label{fig:nmr-history}
\end{figure}

\begin{figure}[H]
\centering
\begin{tikzpicture}[node distance=3cm]
\tikzstyle{tx}=[draw=none, fill=none]
\node[tx] (T1) at (0,0) {$T_1$};
\node[tx] (T2) at (2,0) {$T_2$};

\node[tx] (T3) at (0,-1) {$T_3$};
\node[tx] (T4) at (2,-1) {$T_4$};

\draw[->] (T1) edge node[above]{$ww_x$} (T2);
\draw[->] (T3) edge node[below]{$s_i$} (T4);

\draw[->] (T2) edge node[sloped, above]{$wr_x$} (T3);
\draw[dashed, ->] (T4) edge [bend right]node[right]{$rw_x$} (T2);
\end{tikzpicture}
\caption{DSG for Figure~\ref{fig:nmr-history}. $wr_x$ dependency from $T_1$ to $T_4$ omitted.} 
\label{fig:nmr-dsg}
\end{figure}
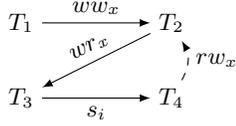

\begin{definition}[Non-monotonic Writes (N-MW)]
A history $H$ exhibits phenomenon N-MW if $DSG(H)$ contains a directed cycle
consisting of a transitive session-dependency between transactions
$T_j$ and $T_i$ and at least one write-dependency edge.
\end{definition}

\begin{figure}[H]
\begin{align*}
\small
T_1 &: w_x(1)\\
T_2 &: w_y(1)\\
T_3 &: r_y(1)~r_x(0)
\end{align*}
\caption{Example of \textit{N-MW} anomaly if $T_2$ directly session-depends on $T_1$.}
\label{fig:nmw-history}
\end{figure}

\begin{figure}[H]
\centering
\begin{tikzpicture}[node distance=3cm]
\tikzstyle{tx}=[draw=none, fill=none]
\node[tx] (T1) at (0,0) {$T_1$};
\node[tx] (T2) at (2,0) {$T_2$};
\node[tx] (T3) at (2,-1) {$T_3$};

\draw[->] (T1) edge node[above]{$s_i$} (T2);
\draw[->] (T2) edge node[right]{$wr_y$} (T3);
\draw[->, dashed] (T3) edge node[below, sloped]{$rw_x$} (T1);
\end{tikzpicture}
\caption{DSG for Figure~\ref{fig:nmw-history}.}
\label{fig:nmw-dsg}
\end{figure}
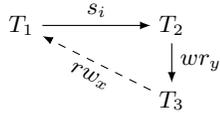

\begin{definition}[Monotonic Writes (MW)]
A system provides Monotonic Writes if it prohibits phenomenon N-MW.
\end{definition}

\begin{definition}[Missing Read-Write Dependency (MRWD)]
A history $H$ exhibits phenomenon MRWD if, in $DSG(H)$, for all
committed transactions $T_1$, $T_2$, $T_3$ such that $T_2$
write-depends on $T_1$ and $T_3$ write-depends on $T_2$, $T_3$ does
not directly anti-depend on $T_1$.
\end{definition}

\begin{figure}[H]
\begin{align*}
\small
T_1 &: w_x(1)\\
T_2 &: r_x(1) w_y(1)\\
T_3 &: r_y(1)~r_x(0)
\end{align*}
\caption{Example of \textit{MRWD} anomaly.}
\label{fig:nwfr-history}
\end{figure}

\begin{figure}[H]
\centering
\begin{tikzpicture}[node distance=3cm]
\tikzstyle{tx}=[draw=none, fill=none]
\node[tx] (T1) at (0,0) {$T_1$};
\node[tx] (T2) at (2,0) {$T_2$};
\node[tx] (T3) at (2,-1) {$T_3$};

\draw[->] (T1) edge node[above]{$wr_x$} (T2);
\draw[->] (T2) edge node[right]{$wr_y$} (T3);
\draw[->, dashed] (T3) edge node[below, sloped]{$rw_x$} (T1);
\end{tikzpicture}
\caption{DSG for Figure~\ref{fig:nwfr-history}.}
\label{fig:nwfr-dsg}
\end{figure}

\begin{definition}[Writes Follow Reads (WFR)]
A system provides Writes Follow Reads if it prohibits phenomenon MWRD.
\end{definition}

\begin{definition}[Missing Your Writes (MYR)]
A history $H$ exhibits phenomenon MYR if $DSG(H)$ contains a directed cycle
consisting of a transitive session-dependency between transactions
$T_j$ and $T_i$, at least one anti-dependency edge, and the remainder
anti-dependency or write-dependency edges.
\end{definition}

\begin{figure}[H]
\begin{align*}
\small
T_1 &: w_x(1)\\
T_2 &: r_x(0)\\
\end{align*}
\caption{Example of \textit{MYR} anomaly if $T_2$ directly session-depends on $T_1$.}
\label{fig:nryw-history}
\end{figure}

\begin{figure}[H]
\centering
\begin{tikzpicture}[node distance=3cm]
\tikzstyle{tx}=[draw=none, fill=none]
\node[tx] (T1) at (0,0) {$T_1$};
\node[tx] (T2) at (2,0) {$T_2$};

\draw[->] (T1) edge node[above]{$s_i$} (T2);
\draw[dashed, ->] (T2) edge [bend left] node[below]{$rw_x$} (T1);
\end{tikzpicture}
\caption{DSG for Figure~\ref{fig:nwfr-history}.}
\label{fig:nryw-dsg}
\end{figure}

\begin{definition}[Read Your Writes (RYW)]
A system provides Read Your Writes if it prohibits phenomenon MYR.
\end{definition}

\begin{definition}[PRAM Consistency]
A system provides PRAM Consistency if it prohibits phenomenon N-MR,
N-MW, and MYR.
\end{definition}

\begin{definition}[Causal Consistency]
A system provides Causal Consistency if it provides PRAM Consistency
and prohibits phenomenon MWRD.
\end{definition}

\begin{definition}[Lost Update]
A history $H$ exhibits phenomenon Lost if $DSG(H)$ contains a directed
cycle having one or more item-antidependency edges and all edges are
by the same data item $x$.
\label{def:lostupdate}
\end{definition}

\begin{definition}[Write Skew (Adya G2-item)]
A history $H$ exhibits phenomenon Write Skew if $DSG(H)$ contains a directed
cycle having one or more item-antidependency edges.
\end{definition}

For Snapshot Isolation, we depart from Adya's recency-based definition
(see Adya Section 4.3). Nonetheless, implementations of this
definition will still be unavailable due to reliance of preventing
Lost Update.

\begin{definition}[Snapshot Isolation]
A system that provides Snapshot Isolation prevents phenomena G0, G1a,
G1b, G1c, PMP, OTV, and Lost Update.
\end{definition}

For Repeatable Read, we return to Adya.

\begin{definition}[Repeatable Read]
A system that provides Repeatable Read Isolation prohibits phenomena
G0, G1a, G1b, G1c, and Write Skew.
\end{definition}

For definitions of safe, regular, and linearizable register semantics,
we refer the reader to Herlihy's textbook~\cite{herlihy-art}.

\section{Monotonic Atomic View Isolation}

In this section, we provide additional information on our two-phase
algorithm for providing Monotonic Atomic View isolation. This section
is effectively a work-in-progress; we are actively investgating
alternative algorithms (particularly with respect to metadata
overheads) and stronger (but still HAT) semantics that complement the
techniques presented here. The interested reader (or would-be systems
implementer) should contact Peter Bailis for more details.

The below pseudocode describes a straightforward implementation of the
algorithm from Section~\ref{sec:ta}. Replicas maintain two sets of
data items: \texttt{pending} (which contains writes that are not yet
pending stable) and \texttt{good} (which contains writes that are
pending stable). Incoming writes are added to \texttt{pending} and
moved to \texttt{good} once they are stable. Stable pending
calculation is performed by counting the number of acknowledgments
(\texttt{acks}) of writes from other replicas, signifying their
receipt of the transaction's writes in their
own \texttt{pending}. Accordingly, the algorithm maintains the
invariant that the transactional siblings (or suitable replacements)
for every write in \texttt{good} are present on their respective
replicas. Clients use the RC algorithm from
Section~\ref{sec:isolation} and keep a map of data items to timestamps
(\texttt{required})---effectively a vector clock whose entries are
data items (not processes, as is standard).

\vspace{1em}\noindent\textbf{Example.} To illustrate this algorithm consider an execution of the following transactions:
\begin{align*}
\small\vspace{-1em}
T_1 &: w_x(1)~w_y(1)
\\T_2 &: r_x(1)~r_y(a)
\end{align*}
For MAV, the system must ensure that $a=1$. Consider a system with one
replica for each of $x$ and $y$ (denoted $R_x$ and $R_y$). Say client
$c_1$ executes $T_1$ and commits. The client chooses timestamp $10001$
(whereby the last $4$ digits are reserved for client ID to ensure
uniqueness) and sends values to $w_x(1)$ to $R_x$ and $R_y$ with
$tx\_keys = \{x,y\}$ and $timestamp=10001$. $R_x$ and $R_y$ place
their respective writes into their individual \texttt{pending} sets
and send acknowledgment for transaction timestamped $10001$ to the
other. Consider the case where $r_x$ has seen $r_y$'s acknowledgment
but not vice versa: $w_x(1)$ will be in \texttt{good} on $R_x$
but \texttt{pending} on $R_y$. Now, if another client $c_2$ executes
$T_2$ and reads $x$ from $R_x$, it will read $x=1$. $c_2$ will update
its \texttt{required} vector so that $\texttt{required}=\{x:10001,
y:10001\}$. When $c_2$ reads $y$ from $R_y$, it will specify
$ts=10001$ and $R_y$ will not find a suitable write
in \texttt{good}. $R_y$ will instead return $y=1$
from \texttt{pending}.

\vspace{1em}\noindent\textbf{Overheads.} MAV requires overhead both on client and server. First, clients must store a vector (\texttt{required}) for the duration of a transaction in order to determine what values to read. Second, each write contains metadata ($tx\_keys$) linear in the number of writes in the transaction. (Note that we could have stored a different timestamp with every write, but a single timestamp suffices and is more efficient) These overheads are modest for short transaction lengths but they increase value sizes on disk, in memory, and over the network. Third, on the server side, each client \textsc{put} incurs two back-ends \texttt{put}s: one into \texttt{pending} and \texttt{good}. If keeping a single value for \texttt{good}, the second \texttt{put} into \texttt{good} (as is the case with normal, single-valued eventually consistent operation) is really a \texttt{get} then a possible \texttt{put}: the server first checks if a higher-timestamped value has been written and, if not, it overwrites the current value. The \texttt{put} into \texttt{pending} is simpler: if message delivery is exactly-once, then \texttt{put}s can be made unconditionally.

\vspace{1em}\noindent\textbf{Optimizations.} The use of \texttt{pending} and \texttt{required} is effectively to
handle the above race condition, where some but not all replicas have
realized that a write is pending stable. With this in mind, there are
several optimizations we can perform:

If a replica serves a read from \texttt{pending}, it must mean one of
two things: either the write is stable pending or the writing client
for the write is requesting the data item (this second case does not
occur in the below pseudocode since own-writes are served out
of \texttt{write\_buffer}). In the former case, the replica can safely
install the write into \texttt{good}.

As we mention in Section~\ref{sec:ta}, the size of \texttt{good} can
be limited to a single item. Recall that any item can be served
from \texttt{good} as long as its timestamp is greater than the
required timestamp. Accordingly, if replicas store multiple writes
in \texttt{good}, any time they can serve a write from \texttt{good},
the highest-timestamped write is always a valid response candidate
(and, indeed, returning the highest-timestamped write provides the
best visibility). Note, however, that serving a higher-than-requested
timestamp from \texttt{pending} is dangerous: doing so might mean
returning a write that is not pending stable and, via appropriate
unfortunate partition behavior, force a client to stall while an
available replica waits for the write to arrive in \texttt{pending}.

The below pseudocode implements a push-based acknowledgment scheme: replicas
eagerly notify other replicas about \texttt{pending} writes. This need
not be the case and, indeed, it is wasteful: if a write
in \texttt{pending} has lower timestamp than the highest timestamped
write in \texttt{good}, it can safely be discarded. Especially for
data items with high temporal locality or data stores that are
replicated over WAN, this \texttt{pending} invalidation may be a
frequent event. Instead, replicas can periodically poll other replicas
for acknowledgments (i.e., asking for either a match
in \texttt{pending} or a higher timestamp in \texttt{good}); this
reduces both network utilization and, in the case of
durable \texttt{pending} and \texttt{good} sets, disk utilization.

There are ample opportunities for batching of pending and good
acknowledgments. Batching is easier in a push-based acknowledgment
scheme, as pull-based acknowledgments may require invalidation to
actually reduce network transfers. Moreover, if a replica is
responsible for several writes within a transaction, \textrm{notify}
messages can be coalesced. We have also observed that batching
multiple background operations together (particularly for
anti-entropy) allows for more efficient compression---especially
important in network-constrained environments.

\vspace{1em}\noindent\textbf{Implementation note.} Our implementation from Section~\ref{sec:evaluation} implements the below pseudocode, with batching between \textrm{notify} operations and a single-item \texttt{good}. We have not yet investigated the efficiency of a pull-based approach (mostly due to our choice of RPC library, which unfortunately does not currently handle multiple outstanding requests per socket). We have also not experimentally determined the visibility effect of the side-channel pending stable notifications as in the first optimization.

\begin{algorithm}[t!]
\newcommand{\myindent}{\hspace{-1em}}
\begin{addmargin}[-1em]{0em}

\begin{algorithmic}
\caption*{\textbf{Algorithm} HAT Read Atomic Isolation}\vspace{.5em}
\State{\textit{\textbf{Shared Data Types}}\\\vspace{.5em}
\texttt{timestamp}: unique, per transaction identifier\\
$\texttt{write}:[\texttt{key:}k, \texttt{value}:v, \texttt{timestamp}:ts, \texttt{set<key>}:sibs]$\\\hrulefill}\vspace{.5em}

  \State{\textbf{\textit{Server-side Data Structures and Methods}\\\vspace{.5em}
$\texttt{set<write>:pending}$\\
$\texttt{set<write>:good}$\\
$\texttt{map<timestamp, int>:acks}$}}\\

\Procedure{put}{\texttt{write}:$w$}
  \State \texttt{pending}.add($w$)
  \For{key $k_t \in w.sibs$}
  \State \{all replicas for $sib$\}.notify($w.ts$)
  \EndFor
  
  \State \textit{asynchronously send $w$ to other replicas via anti-entropy}
  \State \Return

  \EndProcedure\vspace{.5em}

  \Procedure{notify}{\texttt{timestamp}:$ts$}
  \State \texttt{acks}.get($ts$).increment()
  \If{all \texttt{acks} received for all replicas for $ts$'s $sibs$}
  \State \texttt{good}.add($w$)
  \State \texttt{pending}.remove($w$)  
  \EndIf  
  \EndProcedure\vspace{.5em}

  \Procedure{get}{\texttt{key}:$k$, \texttt{timestamp}:$ts_{required}$}
  \If{$ts_{required} = \bot$}
  \State \Return $w \in \texttt{good}$ s.t. $w.key = key$ with highest timestamp
  \ElsIf{$\exists~w \in \texttt{good}$ s.t. $w.key = key, w.ts \geq ts_{required}$}
  \State \Return $w$
  \Else
  \State \Return $w \in \texttt{pending}$ s.t. $w.key = key,  w.ts = ts_{required}$
  \EndIf
  \EndProcedure\\\vspace{-.5em}\hrulefill\vspace{.25em}

  \State{\textbf{\textit{Client-side Data Structures and Methods}\\\vspace{.5em}
$\texttt{int:cur\_txn}$\\
$\texttt{map<key, value>:write\_buffer}$\\
$\texttt{map<key, timestamp>:required}$}}\\

\Procedure{begin\_transaction}{}
  \State \texttt{cur\_txn} = \textit{new txn ID} \commentt{// (e.g., clientID+logicalClock)}
  \EndProcedure\vspace{.5em}

\Procedure{put}{\texttt{key:$k$}, \texttt{value:$v$}}
  \State \texttt{write\_buffer}.put($k$, $v$)
\EndProcedure\vspace{.5em}

\Procedure{get}{\texttt{key:$k$}}
  \If {$k \in \texttt{write\_buffer}$}
  \State \Return \texttt{write\_buffer}.get($k$) \commentt{// if we want per-TxN RYW}
  \EndIf

  \State $w\_{ret}$ = (available replica for $k$).get($k$, \texttt{required}.get($k$))
  \For {$(tx_{key} \in w_{ret}.sibs$}
  \If {$w_{ret}.ts > \texttt{required}.get(tx_{key})$}
  \State $\texttt{required}$.put$(tx_{key}, w_{ret}.ts)$
  \EndIf
  \EndFor

  \State \Return $w\_{ret}.value$
\EndProcedure\vspace{.5em}

\Procedure{commit}{}
 \For{$(tx_{key}, v) \in \texttt{write\_buffer}$}
 \State $r$ = (available replica for $tx_{key}$)
 \State $r$.put($[tx_{key}, v, \texttt{cur\_txn}, \texttt{write\_buffer}.keys()]$)
 \EndFor
  \State \textsc{cleanup()}
\EndProcedure\vspace{.5em}

\Procedure{abort}{}
  \State \textsc{cleanup()}
\EndProcedure\vspace{.5em}

\Procedure{cleanup}{}
  \State \texttt{write\_buffer}.clear()
  \State \texttt{required\_map}.clear()
\EndProcedure\vspace{.5em}

\end{algorithmic}
\end{addmargin}
\end{algorithm}

\end{appendix}
\fi
\end{document}